\documentclass[11pt,a4paper]{article}
\usepackage{amsmath,dsfont}
\usepackage{comment}
\usepackage{wrapfig}
\newcommand{\be}{\begin{equation}}
\newcommand{\ee}{\end{equation}}
\newcommand{\sech}{\mathrm{sech}}

\newcommand{\B}{\mathcal{B}}
\newcommand{\Bb}{\mathrm{B}}
\newcommand{\nn}{\nonumber}
\newcommand{\picco}[1]{\mbox{\small $\displaystyle{#1}$}}
\newcommand{\PP}{\mathrm{P}}
\usepackage{graphicx}
\usepackage{jheppub}
\usepackage[utf8]{inputenc}
\usepackage{physics} 
\usepackage{bbold}
\usepackage{titlesec}
\usepackage{amsfonts}
\usepackage{amssymb}

\author[a]{Luca Griguolo,}
\author[a]{Jacopo Papalini,}
\author[b]{and Domenico Seminara}
 
\affiliation[a]{Dipartimento SMFI, Universit\`a di Parma and INFN Gruppo Collegato di Parma, Viale G.P. Usberti 7/A, 43100 Parma, Italy}
\affiliation[b]{Dipartimento di Fisica, Universit\`a di Firenze and INFN Sezione di Firenze, via G. Sansone 1, 50019 Sesto Fiorentino, Italy\\} 
\emailAdd{luca.griguolo@unipr.it} 
\emailAdd{jacopo.papalini@unipr.it} 
\emailAdd{seminara@fi.infn.it}

\title{On the perturbative expansion of exact bi-local correlators in JT gravity}

\abstract{We study the perturbative series associated to bi-local correlators in Jackiw-Teitelboim (JT) gravity, for positive weight $\lambda$ of the matter CFT operators. Starting from the known exact expression, derived by CFT and gauge theoretical methods, we reproduce the Schwarzian semiclassical expansion beyond leading order. The computation is done for arbitrary temperature and finite boundary distances, in the case of disk and trumpet topologies. A formula presenting the perturbative result (for $\lambda \in \mathbb{N}/2$) at any given order in terms of generalized Apostol-Bernoulli polynomials is also obtained. The limit of zero temperature is  then considered, obtaining a compact expression that allows to discuss the asymptotic behaviour of the perturbative series. Finally we highlight the possibility to express the exact result as particular combinations of Mordell integrals.}
\begin{document}
\maketitle

\pagebreak
\section{Introduction}
Understanding the quantum behavior of gravitational theories is one of the most fascinating problems in contemporary theoretical physics. Jackiw-Teitelboim (JT) gravity \cite{Jackiw:1984je,Teitelboim:1983ux} is probably the simplest example where many questions concerning the nature of a quantum space-time can be tackled and answered, being solvable as a quantum field theory but still retaining quite non-trivial dynamics. It represents a particular example of AdS/CFT correspondence in which we can study bulk and boundary properties with high precision \cite{Almheiri:2014cka,Jensen:2016pah,Maldacena:2016upp,Engelsoy:2016xyb}. Recent investigations have shown that a detailed knowledge of the quantum theory is compulsory to obtain an adequate understanding of quantum gravity's fundamental questions. In particular the physical interpretation of different topological contributions to the gravitational path-integral \cite{Saad:2019lba,Saad:2019pqd} and the advances on the black hole information paradox \cite{Saad:2019pqd,Almheiri:2019qdq,Penington:2019kki} in this setting heavily relied on explicit quantum results. In this sense, JT gravity has played an almost unique role, and it would be crucial to extend our control to more complicated models, generalizing safely some important lessons learned there \cite{Witten:2020wvy,Maxfield:2020ale,Maloney:2020nni,Afkhami-Jeddi:2020ezh,Cotler:2020ugk}. An appealing aspect of JT gravity is the existence of a particular class of $n$-point functions that we can compute exactly, the so-called bi-local correlators \cite{Maldacena:2016upp}. They can be viewed either as $n$-point vacuum expectation value for bi-local operators evaluated at the boundary of the AdS$_2$ space-time or as 2$n$-point correlators of some 1D 'matter CFT' at finite temperature coupled to the Schwarzian theory on the boundary \cite{Maldacena:2016upp}. Their general structure on disk and trumpet topologies has been studied by exploiting different techniques, and their explicit form can be systematically obtained for any $n$ as an integral of momentum space amplitudes, using a simple set of diagrammatic rules. Originally the derivation relied on the precise equivalence between the 1D Schwarzian theory and a certain large central charge limit of 2D Virasoro CFT \cite{Mertens:2017mtv}. More recently, taking advantage of the $\mathrm{SL}(2,{\mathbb R})$ gauge theory formulation (see \cite{Ferrari:2020yon} for an exhaustive analysis of the subject), the correlation functions of bi-local operators have been computed as correlators of Wilson lines anchored at two points on the boundary \cite{Iliesiu:2019xuh}. The time ordering is encoded into the intersection of the Wilson lines in the bulk, resulting in the appearance of momentum-dependent fusion coefficients and 6-$j$ symbols inside the integrated amplitudes. Anchored Wilson lines also have a gravitational interpretation, representing the sum over all possible world-line paths for a particle moving between two fixed points on the boundary of the AdS$_2$ patch \cite{Iliesiu:2019xuh,Blommaert:2018oro}. The computation of bi-local correlators have been later extended in the presence of defects \cite{Mertens:2019tcm}, and the inclusion of higher-genus corrections was also considered \cite{Saad:2019pqd}, with particular attention to their late time behavior and non-perturbative properties. On the other hand, correlation functions on the disk can also be studied through a perturbative expansion in the Schwarzian coupling constant \cite{Maldacena:2016upp,Sarosi:2017ykf}. In this approach, one directly computes Feynman diagrams for boundary gravitons, i.e., the quantum mechanical degrees of freedom associated with the fluctuations of the wiggle AdS$_2$ boundary. The semiclassical limit and the first quantum correction to two-point and four-point functions were studied in \cite{Lam:2018pvp}. Schwarzian perturbation theory has also found applications for higher-point functions \cite{Haehl:2017pak}, while higher loop corrections were analyzed in \cite{Qi:2019gny}. 

Quite surprisingly, the consistency of the exact results obtained through CFT and gauge theoretical techniques with the Schwarzian perturbative expressions has never been checked or diskussed in details until recently\footnote{The semiclassical limit for the two-point and the four-point functions has been checked in \cite{Lam:2018pvp}} \cite{Mertens:2020pfe}. More generally, the structure of the perturbative series and its convergence properties have been somehow overlooked despite certain interesting pieces of information that could be directly extracted from it, as the relation with the gravitational S-matrix or the trigger of the full quantum regime at large time with respect to Schwarzian coupling constant. A particularly intriguing point concerns the convergence itself of the perturbative series and the presence of non-perturbative contributions inside the exact expressions derived in \cite{Mertens:2017mtv,Iliesiu:2019xuh}. A first attempt to answer this question has been taken in \cite{Mertens:2020pfe}: the exact two-point correlator on the disk for a bi-local operator of conformal weight $\lambda\in -{\mathbb N}/2$ have been expanded for a small value of $\kappa$, the  Schwarzian coupling, and confronted successfully with the perturbative result beyond the semiclassical regime. Moreover, exploiting the simplicity of the cases $\lambda=-1/2$ and $\lambda=-1$ and the limit of zero temperature, it was argued that for generic $\lambda$ the series is asymptotic, implying the presence of non-perturbative contributions. Exactly for $\lambda\in -{\mathbb N}/2$ the series appears instead convergent: this particular class of two-point functions can be obtained from a double-scaling limit on degenerate correlators in the Liouville minimal string \cite{Saad:2019lba,Mertens:2020hbs}, suggesting the existence of an integrable subsector in JT gravity. On the contrary, the asymptotic character for generic conformal weights was taken as a signal of non-perturbative contributions of order $e^{-\frac{1}{\kappa}}$ inside these correlators, competing therefore with the higher-genus corrections of order $e^{-\frac{1}{G_N}}$, derived by matrix-model techniques \cite{Saad:2019lba,Saad:2019pqd}, because $\kappa$ is proportional to  gravitational Newton constant $G_N$.

In this paper, we obtain some progress in these directions, performing explicit computations and elucidating the analytical structure 
of the bi-local correlators in the case of general positive conformal weight and general temperature. We also slightly extend our analysis beyond the disk topology by considering the trumpet configuration: this could be relevant in view of further studies on higher-genus topologies \cite{Mertens:2020pfe,Mertens:2020hbs} or for investigating one-point functions in the presence of defects \cite{Mertens:2019tcm}. Our first aim is to recover the Schwarzian perturbative result beyond the leading semiclassical order in the case of positive $\lambda$, on the disk and trumpet topologies. We have obtained a perfect agreement by evaluating the exact expression through a saddle-point approximation: the computation heavily relies on the relevant amplitudes' analytical properties. In the general case, it reduces to an integral around branch-cuts determined by the conformal weight of the operators involved. The result is obtained for finite boundary separations; it exhibits the correct time periodicity and, as expected in this case, the bi-local correlator is singular at coincident points. The outcome completes and generalizes the analysis of \cite{Mertens:2020pfe}, performed for negative semi-integer weights $\lambda$ and in the particular case of zero temperature, and strengths our trust in the analytical approach. Actually, in the case of $\lambda\in {\mathbb N}/2$ we can go well beyond the first subleading quantum correction; the branch-cut singularity of the two-point function reduces to a pole, and by carefully computing the residue, we obtain an all-order expansion in the Schwarzian coupling constant $\kappa$. Finally we have also examined the zero-temperature case, in order to recover the results of \cite{Mertens:2020pfe} in this limit: although being potentially singular, as seen from the previous expansions, we have obtained a nice and compact expression in terms of Bernoulli polynomials, consistent with the general result. We can draw from this limit some conclusions on the convergence properties of the perturbative series, confirming its asymptotic character for positive semi-integer weights. Moreover, the alternate sign of the perturbative orders points towards a possible Borel summability for the full series. As a final observation, we point out that the exact expression for the bi-local correlator can also be written in terms of Mordell integrals \cite{mordell1933}, suggesting a link with the world of Mock-modular forms \cite{2008arXiv0807.4834Z,chen}.

The paper's structure is the following: in Section 2, we review the exact results \cite{Mertens:2017mtv,Iliesiu:2019xuh} and diskuss the perturbative computation of the bi-local correlator in JT gravity from the Schwarzian perspective. We present the explicit calculation on the trumpet to elucidate the procedure, slightly generalizing the previous calculations. In Section 3, we perform the saddle-point analysis of the exact expressions on the disk and the trumpet, successfully recovering the first subleading correction to the semiclassical result. Section 4 is devoted to the all-order expansion in the case $\lambda\in{\mathbb N}/2$. We give the explicit (although a little cumbersome) general expression and examine in more detail the weights $\lambda=1/2$ and $\lambda=1$. The zero-temperature limit is instead the subject of Section 5. Subsequently, in Section 6, we illustrate how the bi-local correlators for integer $2\lambda$ can be written in closed form in terms of Mordell integrals. In Section 7, we draw our conclusions and diskuss some possible directions to extend the present analysis. A certain number of Appendices deepen the technical aspects of the work, completing the paper.

\section{Bi-local correlators in JT gravity: the disk and the trumpet}
\label{bi-local correlators}
This section briefly summarizes how to write down the exact formula for the bi-local boundary correlators in JT gravity, starting from the BF-like picture presented in \cite{Iliesiu:2019xuh}. Here, we will also introduce most of the conventions that we will use later.

\paragraph{Partition function} JT gravity was rephrased in \cite{Iliesiu:2019xuh} as a BF theory with gauge group\footnote{Actually the true gauge group is a certain central extension of $\mathrm{PSL(2,\mathbb{R})}$ by $\mathbb{R}$ \cite{Iliesiu:2019xuh}. For another diskussion of the gauge structure of JT gravity see \cite{Ferrari:2020yon} } $\mathrm{SL(2,\mathbb{R})}$, whose action is 
\begin{equation}\label{BF}
S_{\mathrm{BF}}\left(\phi,\mathcal{A}\right)=-i\int_{\mathcal{M}}\mathrm{Tr}\left(\phi F\right)-\kappa\int_{\partial \mathcal{M}} d \tau \ \mathrm{Tr}\left(\phi^2\right).
\end{equation}
The gauge field $\mathcal{A}$ and the scalar $\phi$ belong to the adjoint representation of $\mathfrak{sl}(2,\mathbb{R})$ algebra. The boundary potential is needed to recover the Schwarzian dynamics on $\partial\mathcal{M}$. We assume that the manifold $\mathcal{M}$ has the topology of the disk. The coupling costant $\kappa$ is related to the gravitational coupling costant $\mathrm{G}_{N}$ by $\kappa=\frac{8 \pi \mathrm{G}_{N}}{\phi_{r}}$ where $\phi_r$ is the renormalized value of dilaton on the boundary.

The structure of the action \eqref{BF} is reminiscent of that of 2D Yang-Mills. However there are two main differences: the quadratic potential is localized on the boundary and the gauge group is non-compact. The partition function for YM$_2$ theory on a disk can be constructed by applying standard Hamiltonian quantization techniques and one gets the following partition function:
\begin{equation}\label{disk}
\mathcal{Z}_{\mathrm{disk}}^{\mathrm{2D} \ \mathrm{YM}}(g,a)=\sum_{\mathrm{R}} \dim \mathrm{R} \ \chi_{\mathrm{R}}(g) \ e^{-\kappa a c_2(\mathrm{R})},
\end{equation}
where the sum extends over all possible representations $\mathrm{R}$ of the compact gauge group, $g$ is the holonomy of $\mathcal{A}$ around the boundary, $a$ the total area of the disk, $c_{2}(\mathrm{R})$ the quadratic Casimir. Finally $\chi_{\mathrm{R}}$ stands for the character basis $\chi_{\mathrm{R}}(g)=\mathrm{Tr}_{\mathrm{R}}(g)$.  

\begin{wrapfigure}{r}{0.52\textwidth}
	\centering
	\includegraphics[width=.7\linewidth]{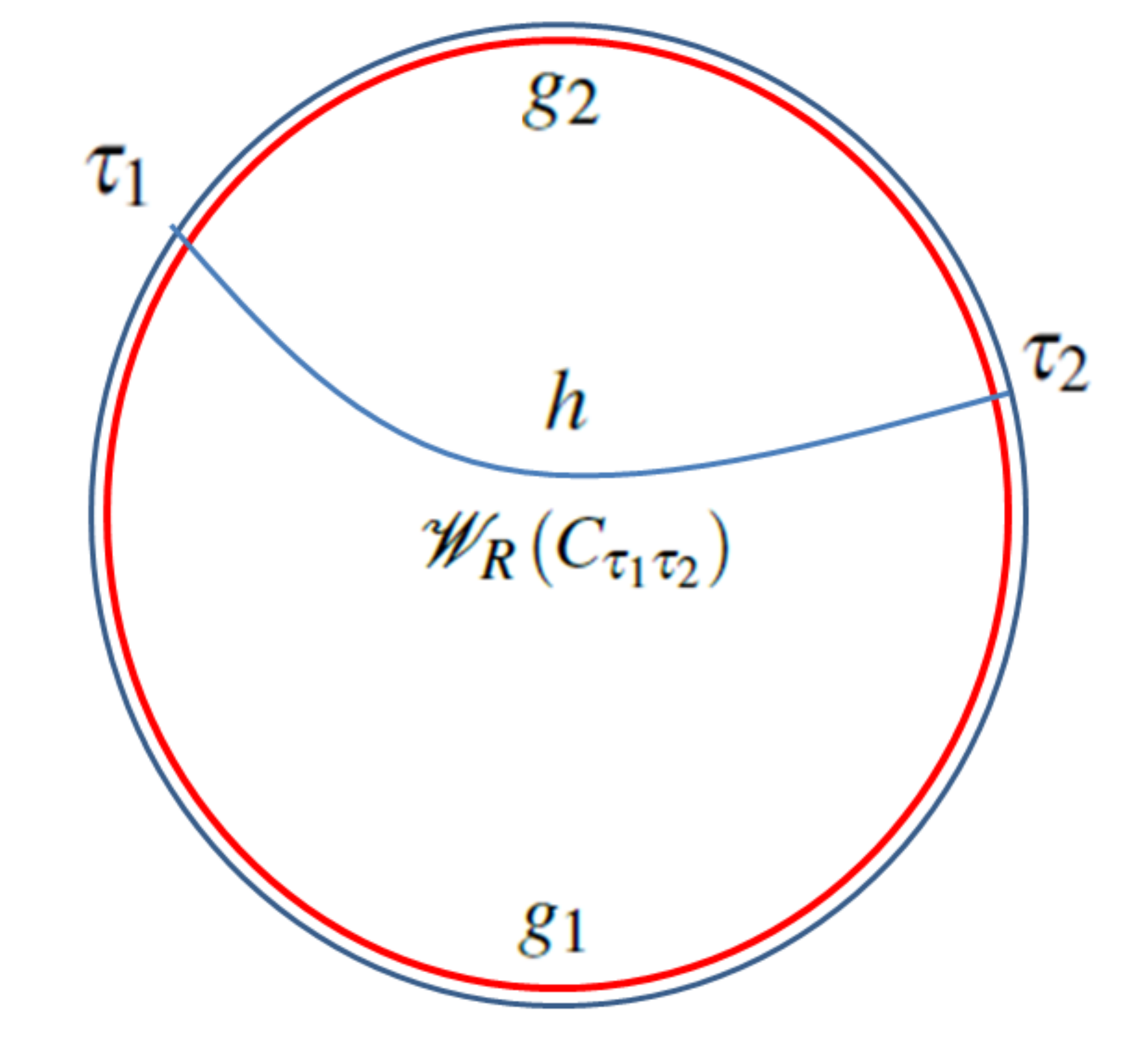}
	\caption{\textit{The blue line crossing the disk corresponds to the boundary anchored Wilson line $\mathcal{W}_{\lambda}\left(\mathrm{C}_{\tau_1 \tau_2}\right)$, which splits the original surface into two regions that can be continuosly deformed into a disk.}}
	\label{fig:wilson line}
\end{wrapfigure} 
 
Following the same logic that led to \eqref{disk}, we find that the disk partition function of the theory \eqref{BF} must have the following structure
\begin{equation}
\mathcal{Z}_{\mathrm{disk}}^{ \mathrm{BF}}(g,\beta)=\int d\mathrm{R} \ \rho(\mathrm{R}) \ \chi_{\mathrm{R}}(g) \ e^{-\kappa \beta c_2(\mathrm{R})}.
\end{equation}
In the case of a non-compact gauge group, such as $\mathrm{SL}\left(2,\mathbb{R}\right)$, the unitary irreducible representations span a continuous spectrum, so we have replaced the sum with an integral, where each representation is weighted with the Plancherel measure $\rho(R)$. Moreover the renormalized boundary length $\beta$ takes the place of the area $a$. For the theory \eqref{BF} one
can argue that only the representations belonging to the Principal series contribute \cite{Iliesiu:2019xuh} and one is left with
\begin{equation}\label{partition}
\mathcal{Z}_{\mathrm{disk}}^{ \mathrm{BF}}(\beta)=\int_{0}^{+\infty} ds \ s \ \sinh \left(2 \pi s\right) e^{-\kappa \beta s^2}=\left(\frac{\pi}{\kappa \beta}\right)^{\frac32} \ e^{\frac{\pi^2}{\beta \kappa}},
\end{equation}
where we have chosen the trivial holonomy around the boundary.
\paragraph{Boundary anchored Wilson lines.} The most interesting and natural observables in the theory \eqref{BF} are Wilson loops or, more
generically, Wilson lines. Since any closed contour is homotopic to a point on the disk, Wilson loops are all trivial for this ({\it almost}) topological theory. On the contrary, Wilson lines whose extrema are anchored to the disk's boundary have a non-trivial expectation value.
This kind of holonomies splits the original disk into two patches, which are homeomorphic to two disks (see fig.~\ref{fig:wilson line}).
Then we can use the {\it cut and sewing} techniques typical of topological theories to compute the expectation value of this observable \cite{Iliesiu:2019xuh}. Specifically we {\it glue}\footnote{Glue operatively means that we integrate over all possible holonomies $h$ on the path $\mathrm{C}_{\tau_1 \tau_2}$ connecting the boundary points $\tau_1$ and $\tau_2$.} together two partition functions of the disk along the common boundary (the blu line $\mathrm{C}_{\tau_1 \tau_2}$ in fig. \ref{fig:wilson line}) where we have also inserted the observable corresponding to the Wilson line, namely
the character $\chi_{\lambda}(h)$ with $h=\mathcal{P} \ \exp \ \int_{C_{\tau_1 \tau_2}} \mathcal{A}$ and $\lambda$ the representation of the Wilson line. We get
\begin{equation}\label{unnorm}
\left< \mathcal{W}_{\lambda}\left(\mathrm{C}_{\tau_1 \tau_2}\right)\right>\propto\int dh \ \mathcal{Z}_{\rm disk}^{\rm BF}\left(g_1h,\tau_{21}\right) \ \chi_{\lambda}(h) \ \mathcal{Z}_{\rm disk}^{\rm BF}\left(h^{-1}g_2,\tau_{12}\right)
\end{equation}
In eq. \eqref{unnorm} $\tau_{12}\equiv \tau$ and $\tau_{21}=\beta-\tau$ are the lengths of the two complementary red boundaries appearing in fig. \ref{fig:wilson line}. Next we substitute the expression for the partition functions of the two disks\footnote{From now on we set the total holonomy around the boundary to be trivial $g_1g_2=1$ and so for semplicity we choose $g_1=g_2=1$.} and we get
\begin{equation}\label{fusion}
\begin{split}
\left< \mathcal{W}_{\lambda}\left(\mathrm{C}_{\tau_1 \tau_2}\right)\right> & \propto \int_{0}^{+\infty} ds_1 \ s_1 \sinh \left(2 \pi s_1\right)e^{-\kappa\tau s_1^2}\int_{0}^{+\infty} ds_2 \ s_2 \sinh \left(2 \pi s_2\right)e^{-\kappa\left(\beta-\tau\right) s_2^2} \ \times \\
& \times \int dh \ \chi_{s_1}(h) \chi_{\lambda}(h) \chi_{s_2}(h^{-1}) 
\end{split}
\end{equation}
The last integral in \eqref{fusion} defines the so called {\it fusion numbers} $\mathcal{N}^{s_2}_{\ s_1,\lambda}$, namely the coefficients
counting how many times the irreducible representation $s_2$ appears into the tensor product of the representations $s_1$ and $\lambda$. For the group $\mathrm{SL}\left(2,\mathbb{R}\right)$, the fusion number is given by a 3-j symbol coefficient
\begin{equation}
\mathcal{N}^{s_2}_{\ s_1,\lambda}=\frac{\Gamma\left[\lambda \pm is_1 \pm is_2\right]}{\Gamma\left[2\lambda\right]}
\end{equation}
where $
\Gamma\left[x\pm y \pm z\right]\equiv\Gamma\left[x+y+z\right]\Gamma\left[x+y-z\right] \Gamma\left[x-y+z\right]\Gamma\left[x-y-z\right]
$.
Thus the exact evaluation of the Wilson line $\left< \mathcal{W}_{\lambda}\left(\tau\right)\right>_{\mathrm{disk}}$ on the disk topology finally yields
\begin{equation}\label{exact}
\left< \mathcal{W}_{\lambda}\left(\tau\right)\right>_{\mathrm{disk}}\!=\!\mathcal{N}_{d}\!\!\int_{0}^{\infty}\!\!\!\!\int_{0}^{\infty} \!\!\!\!\! ds_1ds_2 \ s_1 s_2 \ \sinh\left(2 \pi s_1 \right) \sinh\left(2 \pi s_2 \right) \!\frac{\Gamma\left[\lambda \pm is_1 \pm is_2\right]}{\Gamma\left[2\lambda\right]} \ e^{-\kappa \tau s_1^2 -\kappa \left(\beta -\tau\right)s_2^2}.
\end{equation}
where $\mathcal{N}_{d}$ is a normalization con stant proportional to the inverse of the partition function: we choose it as $\mathcal{N}_{d}\equiv \frac{\kappa^{2\lambda}}{2 \pi}\mathcal{Z}_{\mathrm{disk}}^{-1}=\kappa^{2\lambda}\frac{\left(\kappa \beta\right)^{\frac32}}{2 \pi^{\frac72}}e^{-\frac{\pi^2}{\beta \kappa}}$. The result \eqref{exact} perfectly agrees for instance with the computation performed in \cite{Mertens:2017mtv} via the conformal bootstrap in the Schwarzian theory.

\subsection{The trumpet topology} The first non-trivial topology beyond the disk is the {\it trumpet}, a two dimensional manifold with the topology of the cylinder. The first boundary is analogous to that of the disk, while the other is an asymptotic one whose
geodesic length  is related to the parameter $b$. 
The partition function of JT gravity on this genus-one manifold is given by the integral of a slightly modified spectral density \cite{Mertens:2019tcm}, namely
\begin{equation}\label{trumpet}
\mathcal{Z}_{\mathrm{trump.}}(\beta)=\int_{0}^{+\infty} ds \ \cos\left(2 \pi b s\right) e^{-\kappa \beta s^2}=\left(\frac{\pi}{\kappa \beta}\right)^{\frac12} e^{-\frac{b^2 \pi^2}{\beta \kappa}}
\end{equation}
Following the same logic that led to \eqref{exact} in the case of the disk, we can write down the expectation value of
a boundary anchored Wilson line.  Since this type of path will split the trumpet into two regions, homeomorphic to a disk and a trumpet, 
we can easily show that 
\begin{equation}\label{exact2}
\left< \mathcal{W}_{\lambda}\left(\tau\right)\right>_{\mathrm{trump.}}\!=\!\mathcal{N}_{t}\!\!\int_{0}^{\infty}\!\!\!\!\int_{0}^{\infty} \!\!\!\! ds_1ds_2 s_1  \sinh\left(2 \pi s_1 \right) \cos\left(2 \pi b s_2 \right) \frac{\Gamma\left[\lambda \pm is_1 \pm is_2\right]}{\Gamma\left[2\lambda\right]} \ e^{-\kappa \tau s_1^2 -\kappa \left(\beta -\tau\right)s_2^2}.
\end{equation}
with $\mathcal{N}_{t}\equiv \frac{\kappa^{2\lambda}}{2 \pi}\mathcal{Z}_{\mathrm{trumpet}}^{-1}=\kappa^{2\lambda}\left(\frac{\kappa \beta}{\pi}\right)^{\frac12}e^{\frac{b^2 \pi^2}{\beta \kappa}}$. From the point of view of anchored Wilson loops, this correlator describes bi-local lines not winding around the defects. It was observed in \cite{Mertens:2019tcm} that this observable could not arise from free matter in the bulk
since it does not satisfy the KMS condition (which is equivalent to periodicity around the boundary circle). Therefore they generalized the bi-local operator to satisfy the KMS condition, including an explicit sum over integers in its definition. Computing the correlators with the improved operator is equivalent to sum over Wilson lines encircling the defect, with fixed anchored points. Self-intersections naturally appear for non-trivial windings with the associated 6$j$-symbols, complicating the evaluation of the two-point function.

\subsection{Perturbation theory in the Schwarzian theory}
\label{perturbative computation}
In \cite{Iliesiu:2019xuh} it was suggested that the bulk Wilson line anchored to the points $\tau_1$ and $\tau_2$ of the boundary is dual to the
bi-local correlator of conformal dimension $\lambda$:
\begin{equation}\label{cor}
\mathcal{O}(\tau\equiv \tau_1-\tau_2)=\left[\frac{t'(\tau_1)t'(\tau_2)}{\left(t(\tau_1)-t(\tau_2)\right)^2}\right]^{\lambda}
\end{equation}
computed in the Schwarzian theory, whose action is 
\begin{equation}\label{Sch}
S_{\mathrm{Sch}}\left[t\right]=\frac{\phi_r}{16 \pi \mathrm{G}_{N}}\int_{\partial \mathcal{M}} d \tau \left\{t(\tau),\tau \right\}=\frac{1}{2 \kappa}\int_{\partial \mathcal{M}} d \tau \left\{t(\tau),\tau \right\}.
\end{equation}
Here the fundamental field $t(\tau)$ plays the role of a boundary reparameterization mode, or boundary graviton. The expectation value $\left<\mathcal{O}(\tau)\right>$ is found by inserting \eqref{cor} inside the path integral over the boundary mode $t$ weighted by the Schwarzian action\footnote{ On the disk the Schwarzian path integral is
	\[
	\left<\mathcal{O}(\tau)\right>=\int \frac{\mathcal{D}t}{\mathrm{SL}(2,\mathbb{R})} \ e^{-S_{\mathrm{Sch}}\left[t\right]} \ \mathcal{O}(\tau)
	\]
where $\mathrm{SL}\left(2,\mathbb{R}\right)$ are gauge redundancies of the Schwarzian action. When a hole is inserted, this breaks the gauge group to $\mathrm{U}(1)$.}. We can use this representation to compute these observables perturbatively. This result indirectly provides
a check for the exact formulae \eqref{exact} and \eqref{exact2}. Below, we shall briefly describe how to do this perturbative analysis in the trumpet's less trivial case. For the disk, we refer to \cite{Maldacena:2016upp,Sarosi:2017ykf}. As already remarked in the previous subsection, we do not take into account here the modified definition of the bi-local operator proposed in \cite{Mertens:2019tcm} to implement the KMS condition.

\bigskip

The classical equations of motion for \eqref{Sch} are solved by a field $t(\tau)$ with a constant Schwarzian derivative. In the trumpet case, the classical saddle can be parameterized as
\begin{equation}\label{saddle}
t(\tau)=e^{-\vartheta(\tau)} \qquad \vartheta(\tau)=\frac{2\pi b}{\beta}\left(\tau+\varepsilon (\tau)\right),
\end{equation}
where $\varepsilon(\tau)$ is a small fluctuation over the classical background\footnote{This parametrization can be justified by looking at the metric solution for the disk and the trumpet in Rindler coordinates, which are respectevely 
\[
\mathrm{d}s_{\mathrm{disk}}^2=\mathrm{d}\varrho^2+\sinh^2 \varrho \ \mathrm{d}\tau^2 \qquad \mathrm{d}s_{\mathrm{trumpet}}^2=\mathrm{d}\sigma^2+\cosh^2 \sigma \ \mathrm{d}\vartheta^2 
\]
where the coordinate $\vartheta$ obeys the twisted periodicty $\vartheta \sim \vartheta+b$. The relation between the $\tau$ and $\vartheta$ coordinates at the boundary of the regular hyperbolic disk is $\cos \tau=\tanh \vartheta$ and therefore this implies $t=\tan \frac{\tau}{2}=e^{-\vartheta}$.}. Plugging eq. \eqref{saddle} into eq. \eqref{cor}, we find at zero order in $\varepsilon$  
\begin{equation}\label{tree}
\langle\mathcal{O}(\tau)\rangle_{\rm tree}^{\rm tr.}=\left(\frac{\pi b}{\beta \sinh \left(\frac{\pi b}{\beta}\tau\right)}\right)^{2\lambda},
\end{equation}
which is the tree level amplitude for the correlator on the trumpet geometry. To compute the quantum correction to the tree-level result,
we must determine the propagator for the field $\varepsilon$.
Expanding the action \eqref{Sch} to order $\epsilon^2$ around the saddle \eqref{saddle} we find
\begin{equation}\label{azione}
S_{\varepsilon}=-\frac{\mathrm{1}}{2\kappa}\int_{0}^{\beta} \mathrm{d}\tau \ \left[\varepsilon''(\tau)^2+\left(\frac{2\pi b}{\beta}\right)^2 \varepsilon'(\tau)^2\right]=-\frac{\beta}{2\kappa} \left(\frac{2\pi}{\beta}\right)^4\sum_{n \in \mathbb{Z}} \varepsilon_n \varepsilon_{-n} \ n^2 \left(n^2+b^2 \right),
\end{equation}
where we have Fourier-expanded the fluctuation as $\varepsilon(\tau)=\sum_{n \in \mathbb{Z}}\varepsilon_{n} \ e^{\frac{2\pi i n \tau}{\beta}}$. We recognize the presence of a zero mode $(n=0)$ associated with the residual $\mathrm{U}(1)$ gauge redundancy present in the trumpet geometry. The propagator can be found by inverting the quadratic action and we get
\begin{align}
\label{propa}
\left<\varepsilon(0)\varepsilon(\tau)\right>=&\frac{\kappa\beta^3}{8 \pi^4}\sum_{n \neq 0} \frac{e^{\frac{2\pi i n \tau}{\beta}}}{n^2\left(n^2+b^2 \right)}=\\
=&\frac{\beta  \kappa  \left(\pi ^2 b^2 \left(\beta ^2-6 \beta  \tau +6 \tau ^2\right)+3 \beta ^2-3 \pi  \beta ^2 b \ \text{csch}(\pi  b) \cosh \left(\frac{\pi  b (\beta -2 \tau )}{\beta }\right)\right)}{24 \pi ^4 b^4}.\nn
\end{align}
where the sum over negative and positive integers has been computed in terms of elementary functions by exploiting standard complex analysis techniques \cite{Sarosi:2017ykf}.

\bigskip

To obtain the correction of order $\kappa$ to this observable, we do not need to proceed further in expanding the action. We have instead to expand the bi-local correlator \eqref{cor} around the saddle \eqref{saddle} up to order $\varepsilon^2$. The $\mathcal{O}(\varepsilon)$ has vanishing expectation value since the one-point function is zero for the quadratic action \eqref{azione}. 
Normalizing with respect to the tree level \eqref{tree}, we get
\begin{align}
\frac{\lambda}{2\beta^2} & \left\{4b^2 \pi^2 \left(\lambda \coth^2 \frac{\pi b \tau}{\beta}+\frac12 \mathrm{csch}^2 \frac{\pi b \tau}{\beta}\right)\left(\varepsilon(\tau_1)-\varepsilon(\tau_2)\right)^2+\beta^2\left[\lambda\left(\varepsilon'(\tau_1)+\varepsilon'(\tau_2)\right)^2-\right. \right.\nonumber \\ 
&\left.\left.
-\varepsilon'(\tau_1)^2-\varepsilon'(\tau_2)^2\right] +4\pi b\lambda \beta \coth \left(\frac{\pi b \tau}{\beta}\right)\left[\left(\varepsilon(\tau_2)-\varepsilon(\tau_1)\right)\left(\varepsilon'(\tau_1)+\varepsilon'(\tau_2)\right)\right]\right\}
\end{align}
We now substitute every appearance of $\varepsilon^2-$combination with their expectation value at this order, i.e. with the propagator \eqref{propa} $\left<\varepsilon(0)\varepsilon(\tau)\right>\equiv \mathcal{G}(\tau)$ or its derivatives.
Introducing the auxiliary combination $\xi=\frac{\tau}{\beta}$, the first perturbative term finally reads as
\begin{align}
\frac{\langle\mathcal{O}(\tau)\rangle^{\rm tr.}}{\langle\mathcal{O}(\tau)\rangle_{\rm tree}^{\rm tr.}}&=1+ \frac{\beta  \kappa  \lambda }{4 \pi ^2 b^2} \text{csch}^2(\pi  b \xi )\left[2 \lambda -1-2 \pi ^2 b^2 (\lambda +1) (\xi -1) \xi + \right.\\
&\!\!\!\!\!\!\!\!\!\!\!\!\left.+b\pi (\lambda  (4 \xi -2)-1) \sinh (2 \pi  b \xi )+\left(1-2 \lambda  \left(\pi ^2 b^2 (\xi -1) \xi +1\right)\right) \cosh (2 \pi  b \xi )\right]+O(\kappa^2).\nn
\end{align}
We will reproduce the above expression from the exact formula \eqref{exact2} in the next section.

\section{Recovering the perturbative expansion}
\subsection{Bi-local correlator on the disk}
In the following our goal is to illustrate how the perturbative results for the bi-local correlator $\langle\mathcal{O}^{(\lambda)}(\tau)\rangle_\beta^{\rm disk}$ on the disk can be recovered from its exact integral representation \eqref{exact}. 
A simple-minded Taylor-expansion of the integrand would lead to divergent expressions since the representation \eqref{exact} is naturally suited to derive the large $\kappa$ expansion for the two-point function.

\begin{figure}  
	\centering
	\includegraphics[width=0.9\linewidth]{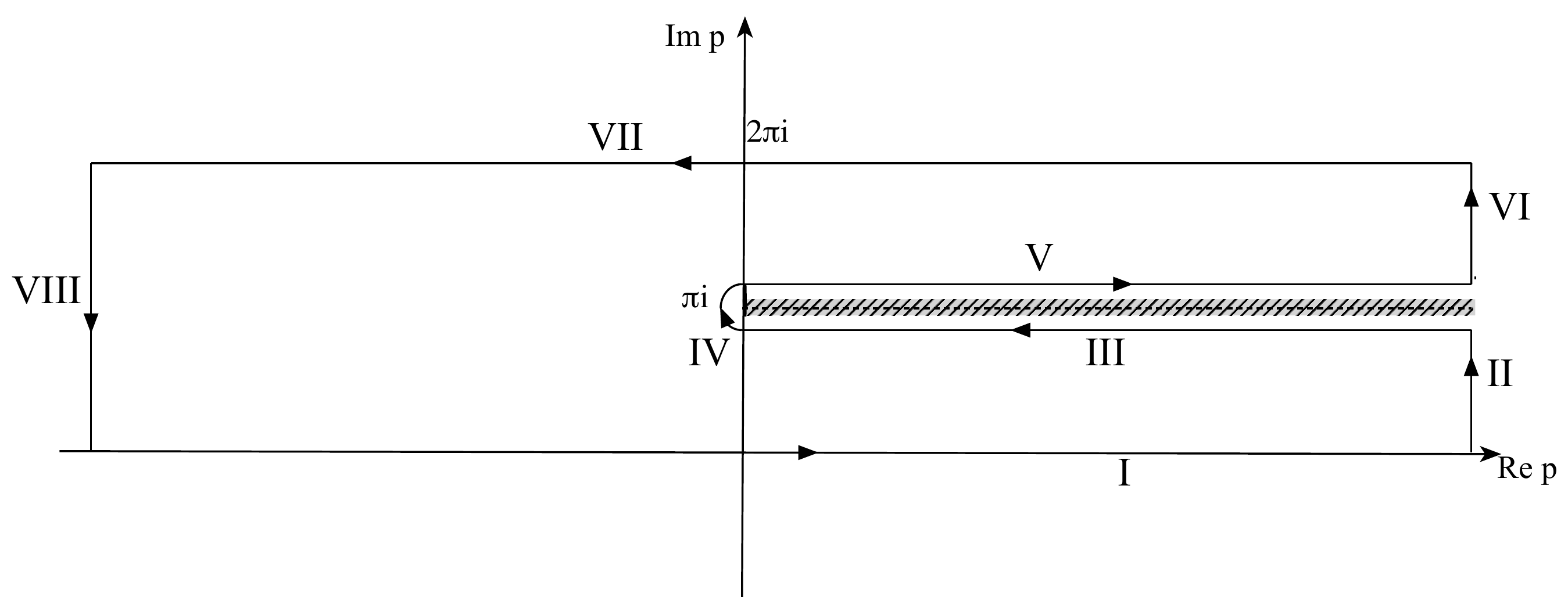}
	\caption{Contour in the complex p-plane used to perform the shift $p\rightarrow p+2\pi i$.}
	\label{fig:cut}
\end{figure}

To obtain the perturbative series in $\kappa$ we have to rearrange the $\kappa$ dependence of \eqref{exact} and we start by using the following identity, first derived by Ramanujan:
\begin{equation}\label{rama}
\begin{split}
\int_{-\infty}^\infty \mathrm{d}p~ \sech^{2 a}\left(\frac{p}{2}\right) e^{i px}=\frac{2^{2a-1}}{\Gamma(2a)} \Gamma\left(a+ i x\right) \Gamma\left(a- i x\right).
\end{split}
\end{equation}
which holds for $\Re(a)>0$. In this way we can express the 3-j symbol in Fourier space and we get
\begin{equation}
\begin{split}\label{bi-local1}
\langle\mathcal{O}^{(\lambda)}&(\tau)\rangle_\beta^{\rm disk}=\frac{\mathcal{N}_d\Gamma(2\lambda)}{2^{4\lambda-2}}\int_{0}^\infty\!\! \int_{0}^\infty \!\!\mathrm{d}s_1 \mathrm{d}s_2 \ s_1 s_2 \ \sinh 2\pi s_1 \sinh 2\pi s_2 \ e^{-\kappa(\beta-\tau) s_1^2 -\kappa \tau s_2^2}
 \times\\
&\times \int_{-\infty}^\infty dp \int_{-\infty}^\infty dq~ {e^{i p(s_1+s_2)+i q(s_1-s_2)}}~\sech^{2\lambda}\frac{p}{2} \sech^{2\lambda}\frac{q}{2}
 \ \end{split}
\end{equation}
Since the integrand is even in $s_1$ and $s_2$, we can extend the region of integration to the entire real line and subsequently
 perform the gaussian integration over $s_1$ and $s_2$. We are left with a double integral over $p$ and $q$:
\begin{equation}
\label{bi-local2}
\langle\mathcal{O}^{(\lambda)}(\tau)\rangle_\beta^{\rm disk}=\frac{\pi\mathcal{N}_d\Gamma(2\lambda)}{2^{4\lambda+2} \kappa ^3 \tau
  ^{3/2} (\beta -\tau )^{3/2}}
 \int_{-\infty}^\infty dp dq~ 
\frac{q^2- (p-2 i \pi )^2}{\cosh^{2\lambda}\frac{p}{2} \cosh^{2\lambda}\frac{q}{2}} e^{-\frac{(p+q-2 i \pi )^2}{4 \kappa 
  (\beta -\tau )}-\frac{(p-q-2 i \pi )^2}{4 \kappa \tau }}
  \end{equation}
The original symmetry in the exchange $\tau\leftrightarrow\beta-\tau$ in \eqref{exact} is now realized by the change of variables
$q\leftrightarrow -q$.
 \noindent Next we perform the shift $p\to p+2\pi i$ by considering the contour displayed in fig. \ref{fig:cut} in the complex $p-$plane. In the following we shall assume that $2\lambda\not \in \mathds{N}$\footnote{The case
 $2\lambda\in \mathds{N}$, where the cut is replaced by a pole, will be diskussed in detail in sec. \ref{allorder}.}. Then the contour encircles the branch cut,
 due to $(\cosh\frac p 2)^{-2\lambda}$, that has been chosen to run from $p=\pi i$ to $p=\infty+\pi i$\footnote{This position of the cut is obtained by choosing the phase around the  branch point between $(-\frac{3\pi}{2}, \frac{\pi}{2}]$.}.
Moreover, we take $0<2\lambda<1$ so that the contribution of the semicircle (IV) is finite and vanishes when we shrink its radius to zero.
At the end we will recover the perturbative result for $2\lambda>1$ by extending analytically the final expression.

\noindent
The contributions of the vertical edges (II, VI and VIII) of the contour vanish  when we approach infinity and thus the original integral (edge I) can be replaced by the two terms
coming respectively from the  horizontal edge (VII) and the diskontinuity around the cut
\begin{equation}
\label{bi-local3}
\begin{split}
\langle\mathcal{O}^{(\lambda)}(\tau)\rangle_\beta^{\rm disk}&=\frac{\pi\mathcal{N}_de^{-2\pi i\lambda}\Gamma(2\lambda)}{2^{4\lambda+2} \kappa ^3 \tau
  ^{3/2} (\beta -\tau )^{3/2}}
 \int_{-\infty}^\infty dp dq~ 
\frac{ q ^2-p^2}{\cosh^{2\lambda}\frac{p}{2} \cosh^{2\lambda}\frac{q}{2}} e^{-\frac{(p+q )^2}{4 \kappa 
  (\beta -\tau )}-\frac{(p-q )^2}{4 \kappa \tau }}-\\
 & -\frac{ \pi i \mathcal{N}_d e^{\pi i\lambda }\sin \left(2\pi \lambda\right)\Gamma(2\lambda)}{2^{4\lambda+1} \kappa ^3 \tau
  ^{\frac{3}{2}} (\beta -\tau )^{\frac{3}{2}}}\!
 \int_{-\infty}^\infty\!\!\!\!\! dq\int_0^\infty\!\!\!\!\! dt~ 
\frac{ q^2-(t- i \pi )^2}{\cosh^{2\lambda}\frac{q}{2} \sinh^{2\lambda}\frac{t}{2}} e^{-\frac{(t+q- i \pi )^2}{4 \kappa 
  (\beta -\tau )}-\frac{(t-q- i \pi )^2}{4 \kappa \tau }}
  \end{split}
  \end{equation}
The first integral in \eqref{bi-local3} vanishes because of the antisymmetry in the exchange $p\leftrightarrow q$. In the second
one we can safely perform the following shift  
\be
q\to q+\frac{(t-i \pi ) (\beta -2 \tau )}{\beta}
\ee
since we do not encounter any branch cut or singularity of the integrand during this process (at least for generic values of $\beta$ and
$\tau$). This shift centers the integral at $q=0$ and we obtain
\begin{align}
\label{bi-local4}
\langle\mathcal{O}^{(\lambda)}(\tau)\rangle_\beta^{\rm disk}=&- \frac{ \pi i \mathcal{N}_d e^{\pi i\lambda }\sin \left(2\pi \lambda\right)\Gamma(2\lambda)}{2^{4\lambda+1} \kappa ^3\beta^3 \xi
  ^{\frac{3}{2}} (1 -\xi )^{\frac{3}{2}}}\!\times\\
&\times \int_{-\infty}^\infty\!\!\!\!\! dq\int_0^\infty\!\!\!\!\! dt~ 
\frac{  (q+2 (t-\pi i ) (1 -\xi )) (q-2( t-\pi i) \xi 
  )}{\cosh^{2\lambda}\frac{q+(t-\pi i)(1-2\xi)}{2} \sinh^{2\lambda}\frac{t}{2}} { e^{-\frac{ q^2}{4  \kappa \beta\xi (1-\xi) 
  }-\frac{(t-\pi i)^2}{\beta \kappa }}}\nonumber
  \end{align}
where we have found it convenient to introduce the auxiliary combination $\xi\equiv\frac{\tau}{\beta}$. The form \eqref{bi-local4} of the integral representation is suited to identify the origin of the dominant contributions in the limit $\kappa\to 0$. A neighborhood around $q=0$ dominates the integration over $q$ due to the integrand's gaussian weight. For the same reason, one might assume that the integration over $t$ is also primarily controlled by a small interval around $t=\pi i$. On the other hand, since $t$ spans the semi-infinite interval $[0,+\infty]$, we have to consider a second candidate, namely the neighbourhood around $t=0$ (see \cite{olver2014asymptotics} for the general theory). Comparing the two possibilities, we find that the integral \eqref{bi-local4} in the limit $\kappa\to0$ is dominated by the second one since the gaussian weight scales as $e^{\frac{\pi^2}{\kappa\beta}}$.

\noindent
The simplest way to construct systematically the asymptotic series in the limit $\kappa\to 0$ is to perform the following rescaling of variables
\be
\label{scale}
 q\mapsto \sqrt{\kappa} q\quad\quad\quad t\mapsto \kappa t.
\ee
The different scaling of the variable $t$ takes into account that the leading contribution comes from the lower extremum of the integral and not from a saddle-point. Using the explicit form of the normalization $\mathcal{N}_d$ we get
\begin{align}
\label{bi-local5}
\langle\mathcal{O}^{(\lambda)}&(\tau)\rangle_\beta^{\rm disk}=-\frac{i e^{  \pi i \lambda } \kappa ^{2 \lambda } \sin (2 \pi 
  \lambda ) \Gamma (2 \lambda )}{2^{4 \lambda +2} \pi ^{5/2} \beta ^{3/2} (1-\xi )^{3/2} \xi ^{3/2}}\!\times\\
&\times \int_{-\infty}^\infty\!\!\!\!\! dq\int_0^\infty\!\!\!\!\! dt~ 
\frac{  (\sqrt{\kappa}q+2 (\kappa t-\pi i ) (1 -\xi )) (\sqrt{\kappa}q-2( \kappa t-\pi i) \xi 
  )}{\cosh^{2\lambda}\frac{\sqrt{\kappa}q+(\kappa t-\pi i)(1-2\xi)}{2} \sinh^{2\lambda}\frac{\kappa t}{2}} { e^{-\frac{ q^2}{4   \beta\xi (1-\xi) 
  }-\frac{\kappa t^2}{\beta }-\frac{2\pi i t}{\beta}}}\nonumber
  \end{align}
The non-analytic factor $e^{-\frac{\pi ^2}{\beta \kappa }}$ present in $\mathcal{N}_d$ cancels exactly against the constant term in the gaussian weight for $t$ and we can Taylor-expand the integrand \eqref{bi-local5} around $\kappa=0$, obtaining a series with both integer and semi-integer powers of $\kappa$. The latter is always proportional to an odd power of $q$ and vanish when the integral is performed. Moreover, the expansion generates integrals over $t$, which are divergent for real $\beta$. We can take care of this issue by rotating our path of integration in $t$ of a small positive angle $\alpha$ before expanding. Once we have integrated over $t$, the final result does not depend on $\alpha.$ Alternatively, we could assume that $\beta$ has a small imaginary part and then analytically continue to real values.

\noindent
After integrating over $t$ and $q$ term by term, we find 
\begin{align}
\label{bi-local6}
&\langle\mathcal{O}^{(\lambda)}(\tau)\rangle_\beta^{\rm disk}=
\frac{\pi ^{2 \lambda } }{\beta ^{2 \lambda } \sin ^{2 \lambda }(\pi \xi )}\Bigl[1+\frac{\kappa\beta \lambda }{4 \pi ^2\sin ^2(\pi \xi )} (2 \pi ^2 (\lambda +1) \xi ^2-2 \pi ^2
  (\lambda +1) \xi -\nonumber\\
  &-\pi (2 \lambda +1) (2 \xi -1) \sin (2 \pi \xi )+(2 \lambda 
  (\pi ^2 (\xi -1) \xi -1)-1) \cos (2 \pi \xi )+2 \lambda
  +1)+O(\kappa^2)\Bigr]
\end{align}
In this expansion, we recognize the classical term and the one-loop contribution obtained by a direct diagrammatic computation in \cite{Sarosi:2017ykf}. A systematic all order expansion can also be obtained by expanding the integrand in terms of generalized Apostol-Eulerian and Bernoulli polynomials. However, the final expression is not particularly appealing, and we will concentrate on the particular case $2\lambda\in \mathbb{N}$. 
\subsection{Bi-local correlator on the trumpet}
The structure of the bi-local operator on the trumpet \eqref{exact2} is quite similar to the case of the disk and if we use the symmetry of the integrand, we can rearrange it in the following form:
\begin{align}
\langle\mathcal{O}^{(\lambda)}&(\tau)\rangle_\beta^{\rm tr.}=\frac{\mathcal{N}_t}{4}\int_{-\infty}^\infty \int_{-\infty}^\infty \mathrm{d}s_2\mathrm{d}s_2 \ s_1 \ e^{2\pi \left(s_1+ibs_2\right)-\kappa(\beta-\tau) s_2^2 -\kappa \tau s_1^2} \times\nonumber\\
&\times \frac{\Gamma\left(\lambda-i s_1-i s_2\right)\Gamma\left(\lambda+i s_1+i s_2\right)
	\Gamma\left(\lambda+i s_1-i s_2\right)\Gamma\left(\lambda-i s_1+i s_2\right)}{\Gamma (2\lambda)}.
\end{align}
As in the case of the disk, we can use the identity \eqref{rama} to eliminate the Gamma function and perform the gaussian integration over
$s_1$ and $s_2$. We find this new integral representation for the bi-local correlator \eqref{exact2}:
\begin{align}
\langle\mathcal{O}^{(\lambda)}(\tau)\rangle_\beta^{\rm tr.}=
\frac{i \pi \mathcal{N}_t \Gamma (2 \lambda ) }{ 2^{4 \lambda +1}\kappa ^2 \sqrt{\tau } (\beta -\tau )^{\frac32}} \int_{-\infty}^\infty\!\!\!\! dp dq~ \frac{(p+q-2\pi i) }{\cosh ^{2
  \lambda }\left(\frac{p}{2}\right) \cosh ^{2 \lambda }\left(\frac{q}{2}\right) }e^{
  -\frac{(2 \pi b+p-q)^2}{4\kappa(\beta -\tau) }-\frac{(p+q-2 i \pi )^2}{4\kappa\tau }}.
\end{align}
Next we shift the variables of integration as follows
$
p\mapsto p-\pi b $ $q\mapsto q+\pi b 
$ and we get
\begin{align}
\langle\mathcal{O}^{(\lambda)}(\tau)\rangle_\beta^{\rm tr.}=
\frac{i \pi \mathcal{N}_t \Gamma (2 \lambda ) }{ 2^{4 \lambda +1}\kappa ^2 \sqrt{\tau } (\beta -\tau )^{\frac32}} \int_{-\infty}^\infty\!\!\!\! dp dq~ \frac{(p+q-2\pi i ) e^{
  -\frac{(p-q)^2}{4\kappa(\beta -\tau) }-\frac{(p+q-2 \pi i )^2}{4\kappa\tau }}}{\cosh ^{2
  \lambda }\left(\frac{p-\pi b}{2}\right) \cosh ^{2 \lambda }\left(\frac{q+\pi b}{2}\right) }.
\end{align}
Again we perform the shift $p\to p+2\pi i$ by considering a contour similar to the one displayed in fig. \ref{fig:cut}. The only difference is the position of the branch cut that now runs $p=\pi b+\pi i$ to $p=\infty+\pi i$. As in the case of the disk we assume that $2\lambda\not \in \mathds{N}$ and $0<2\lambda<1$. We get
\begin{align}
\langle\mathcal{O}^{(\lambda)}&(\tau)\rangle_\beta^{\rm tr.}=
\frac{i \pi e^{-2\pi i\lambda}\mathcal{N}_t \Gamma (2 \lambda ) }{ 2^{4 \lambda +1}\kappa ^2 \sqrt{\tau } (\beta -\tau )^{\frac32}} \int_{-\infty}^\infty\!\!\!\! dp dq~ \frac{(p+q ) e^{
  -\frac{(p-q+2\pi i)^2}{4\kappa(\beta -\tau) }-\frac{(p+q)^2}{4\kappa\tau }}}{\cosh ^{2
  \lambda }\left(\frac{p-\pi b}{2}\right) \cosh ^{2 \lambda }\left(\frac{q+\pi b}{2}\right) }+\\
  &+\frac{
  2 \pi \mathcal{N}_t \Gamma (2 \lambda ) e^{\pi i\lambda }\sin \left(2\pi \lambda\right)}{ 2^{4 \lambda +1}\kappa ^2 \sqrt{\tau } (\beta -\tau )^{\frac32}} \int_{-\infty}^\infty\!\!\!\! dq \int_0^\infty\!\!\!\!\! dt~ \frac{(t+q-\pi i +\pi\beta) e^{
  -\frac{(t-q+\pi b+\pi i)^2}{4\kappa(\beta -\tau) }-\frac{(t+q- \pi i+\pi b )^2}{4\kappa\tau }}}{\sinh ^{2
  \lambda }\left(\frac{t}{2}\right) \cosh ^{2 \lambda }\left(\frac{q+\pi b}{2}\right) }.\nonumber
\end{align}
The first integral vanishes because it is odd under the transformations $p\mapsto -q$ and $q\mapsto -p$. In the second integral
we perform a shift in $q$ to center the gaussian weight around $q=0$ and we obtain the analog of \eqref{bi-local4}:
\begin{align}
\langle\mathcal{O}^{(\lambda)}&(\tau)\rangle_\beta^{\rm tr.}=
\frac{\pi \mathcal{N}_t \Gamma (2 \lambda ) \sin (2\pi \lambda ) }{ 16^{\lambda }\beta ^2 \kappa ^2 \sqrt{1-\xi } \xi
  ^{3/2}} \times\nonumber\\
  &\times\int_{-\infty}^\infty\!\!\!\! dq \int_0^\infty dt~ \frac{2 \xi (\pi b+t)+q}{\sinh ^{2
  \lambda }\left(\frac{t}{2}\right) \sinh^{2\lambda}
  \left(\frac{1}{2} (2 \xi (\pi b+t)+q-t)\right)} e^{
-\frac{(\pi b+t)^2}{\beta \kappa } -\frac{q^2}{4 \beta 
  \kappa \xi (1- \xi )}} \end{align}
 where we have again introduced the auxiliary combination $\xi\equiv\frac{\tau}{\beta}$.  As in the previous case, the integration over $q$ is again dominated by a neighbourhood around $q=0$ due to the gaussian weight in the integrand.  Since $t$ spans the semi-infinite interval $[0,+\infty]$ and in this interval the gaussian weight is monotonic (for $b>0$), we find that  the integral over $t$ in the limit $\kappa\to 0$ is controlled by the lower bound of the integration interval, $t=0$.

Next we scale the variables $t$ and $q$ as in \eqref{scale} and expand the integrand around $\kappa=0$. Performing the two integrations term by term we find
\begin{align}
&\langle\mathcal{O}^{(\lambda)}(\tau)\rangle_\beta^{\rm tr.}=\frac{\pi ^{2 \lambda } b^{2 \lambda }}{ \beta ^{2 \lambda } \sinh ^{2
  \lambda }(\pi b \xi )}
\biggl[1+\frac{\beta \kappa \lambda }{4 \pi ^2 b^2 \text{sinh}^2(\pi b \xi )}
 \left(-2 \pi ^2 b^2 (\lambda
  +1) \xi ^2+2 \pi ^2 b^2 (\lambda +1) \xi +\right.\nonumber\\ &\left.+\!\left(1-2 \lambda \left(\pi ^2 b^2 (\xi
  -1) \xi +1\right)\right)\! \cosh (2 \pi b \xi )\!+\!\pi b (\lambda (4 \xi -2)-1) \sinh
  (2 \pi b \xi )\!+2 \lambda -1\right)\!+\!O(\kappa^2)\biggr]
\end{align}
In this expansion we recognize the classical term and the one-loop contribution obtained by a direct diagrammatic computation in subsec.  
 \ref{perturbative computation}. 
\section{All order expansion of the bi-local correlators: the integer case $2\lambda\in\mathbb{N}$}
\label{allorder}
In this section we focus our attention on a particular but very interesting case, namely $2\lambda\in\mathds{N}$. For semi-integer values of $\lambda$, the cut present in fig. \ref{fig:cut} is replaced by a pole of order $2\lambda$. For this reason it is convenient to start over our analysis from the integral representation \eqref{bi-local2} and use as new variables of integration
\begin{equation} 
p=\frac{u+v}{2} \qquad q=\frac{u-v}{2}.
\end{equation}
We get a nice and symmetric representation for the bi-local correlator on the disk: 
\begin{equation}
\label{bi-localinte2}
\langle\mathcal{O}^{(\lambda)}(\tau)\rangle_\beta^{\rm disk}=-\frac{\pi\mathcal{N}_d\Gamma(2\lambda)}{2^{2\lambda+3} \kappa ^3 \tau
  ^{3/2} (\beta -\tau )^{3/2}}
 \int_{-\infty}^\infty du dv~ 
\frac{(u-2\pi i) (v-2\pi i)}{(\cosh \frac{u}{2}+\cosh \frac{v}{2})^{2\lambda}} e^{-\frac{(u-2 i \pi )^2}{4 \kappa 
  (\beta -\tau )}-\frac{(v-2 i \pi )^2}{4 \kappa \tau }}
  \end{equation}
\begin{figure}[t]
	\centering
	\includegraphics[width=0.9\linewidth]{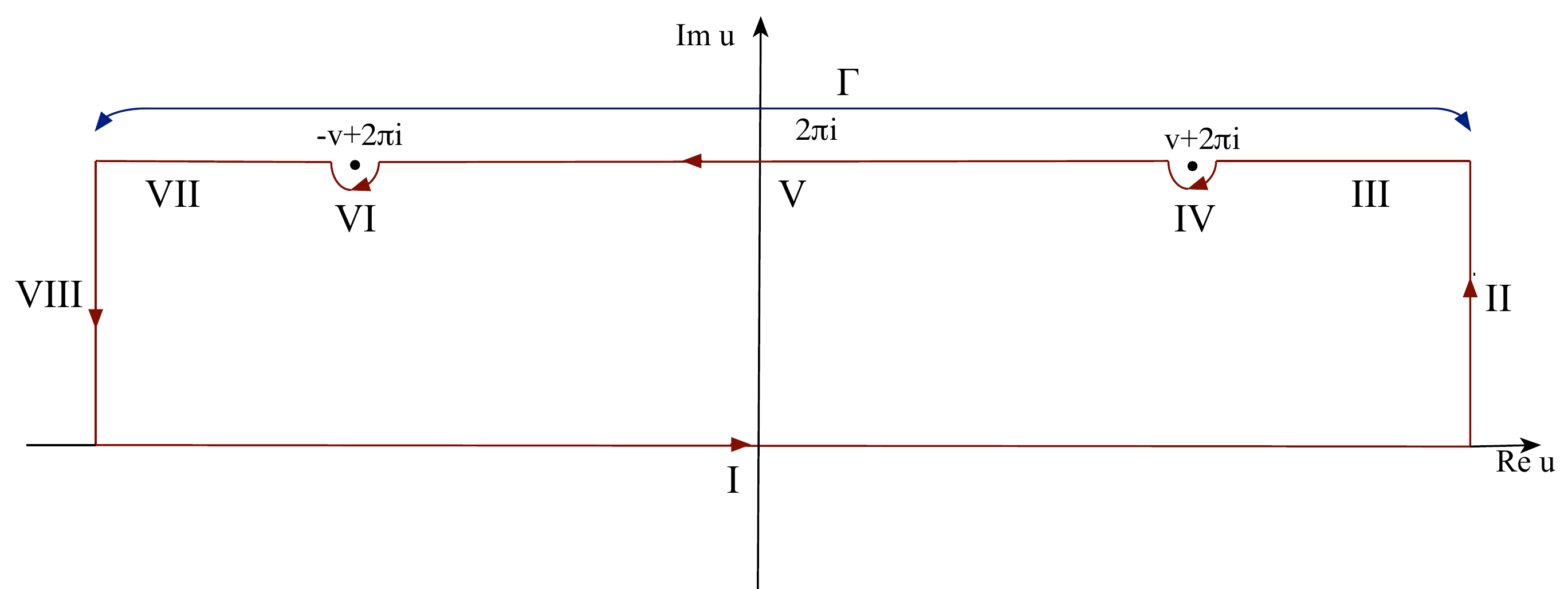}
	\caption{The red contour $C$ used to perform the integration over $u$}
	\label{fig1}
\end{figure}

\noindent
Next we evaluate the integral over $u$ in \eqref{bi-localinte2} using residues. Consider the closed red contour $C$ depicted in fig. \ref{fig1}. Along this path the integral of the function
\begin{equation}
\label{fun}
f(u,v)=-\frac{\pi\mathcal{N}_d\Gamma(2\lambda)}{2^{2\lambda+3} \kappa ^3 \tau
 ^{3/2} (\beta -\tau )^{3/2}}
\frac{(u-2\pi i) (v-2\pi i)}{(\cosh \frac{u}{2}+\cosh \frac{v}{2})^{2\lambda}} e^{-\frac{(u-2 i \pi )^2}{4 \kappa 
 (\beta -\tau )}-\frac{(v-2 i \pi )^2}{4 \kappa \tau }}
\end{equation}
 is identically zero as \eqref{fun} defines a holomorphic function in the enclosed region. Since the contributions of the two vertical edges II and VIII vanish when they approach infinity, the integral along the entire real $u-$axis, i.e. the original integral, is equal to minus the integral of $f(u,v)$ along $\Gamma$ (see fig. \ref{fig1}):
 \begin{equation} 
\begin{split}
	\langle\mathcal{O}^{(\lambda)}(\tau)\rangle_\beta^{\rm disk}=& \frac{\pi\mathcal{N}_d\Gamma(2\lambda)}{2^{2\lambda+3} \kappa ^3 \tau
 ^{3/2} (\beta -\tau )^{3/2}} \int_{-\infty}^\infty\!\!\!\! dv \int_{ \Gamma} \mathrm{d}u \frac{(u-2\pi i)(v-2\pi i)}{\left(\cosh\frac{v}{2}+\cosh\frac{u}{2}\right)^{2\lambda}}e^{-\frac{(v-2 i \pi )^2}{4
			\kappa\tau }-\frac{(u-2 i \pi )^2}{4
			\kappa(\beta -\tau) }}
\end{split}
\end{equation} 
The path $\Gamma$ is composed by three straight segments (III,V and VII) and two semi-circum\-fe\-ren\-ces (IV and VI). The former three contributions either cancel or vanish  because the resulting integrand is an odd function under reflection with respect to the axis Im $u$. Instead the latter two (i.e. IV and VI) yield
 \begin{align} 
 \label{residue}
	&\langle\mathcal{O}^{(\lambda)}(\tau)\rangle_\beta^{\rm disk}=- \frac{\pi^2 i\mathcal{N}_d\Gamma(2\lambda)}{2^{2\lambda+4} \kappa ^3 \tau
 ^{3/2} (\beta -\tau )^{3/2}} \int_{-\infty}^\infty dv~e^{-\frac{(v-2 i \pi )^2}{4
			\kappa\tau } }(v-2\pi i)\times\nonumber\\
 &\times\left( \mathrm{Res}_{u=v+2\pi i}\left[ \frac{(u-2\pi i)e^{-\frac{(u-2 i \pi )^2}{4
			\kappa(\beta -\tau) }}}{\left(\cosh\frac{v}{2}+\cosh\frac{u}{2}\right)^{2\lambda}}\right]+\mathrm{Res}_{u=-v+2\pi i}\left[ \frac{(u-2\pi i)e^{-\frac{(u-2 i \pi )^2}{4
			\kappa(\beta -\tau) }}}{\left(\cosh\frac{v}{2}+\cosh\frac{u}{2}\right)^{2\lambda}}\right]\right)=\\
			&=	- \frac{\pi^2 i\mathcal{N}_d\Gamma(2\lambda)}{2^{2\lambda+3} \kappa ^3 \tau
 ^{3/2} (\beta -\tau )^{3/2}} \int_{-\infty}^\infty dv~(v-2\pi i)e^{-\frac{(v-2 i \pi )^2}{4\kappa\tau}} \mathrm{Res}_{u=v}\left[ \frac{u e^{-\frac{u^2}{4
			\kappa(\beta -\tau) }}}{\left(\cosh\frac{v}{2}-\cosh\frac{u}{2}\right)^{2\lambda}}\right],\nonumber
\end{align}
where we used the symmetry of the integrand to show that the two residues are equal.

\subsection{Some interesting examples: small $\lambda$ values}
The representation \eqref{residue} is very efficient in reconstructing the perturbative series at all orders. To illustrate how we can recover the series for small $\kappa$, we first focus on the case $\lambda=\frac12$. Then the residue in \eqref{residue} can be easily evaluated and is given by
\begin{equation} 
\label{case1}
\mathrm{Res}_{u=v}\left(\frac{u \ e^{-\frac{u^2}{4
				\kappa(\beta-\tau)}}}{\cosh\frac{v}{2}-\cosh\frac{u}{2}} \right)=-2v \ \mathrm{csch} \frac{v}{2} e^{-\frac{v^2}{4 \kappa \left(\beta-\tau\right)}}.
\end{equation}
Next we can recast the integral \eqref{residue} as follows 
\begin{equation} 
\label{case2}
\langle\mathcal{O}^{(\frac12)}(\tau)\rangle_\beta^{\rm disk}=\mbox{\small$ \frac{i\pi^2 \mathcal{N}_de^{\frac{\pi^2}{\beta \kappa}}}{ 2^{2\lambda+1}\kappa ^3 \tau ^{3/2}
		(\beta -\tau )^{3/2}}$} \int_{-\infty}^{+\infty} \mathrm{d}v\frac{v\left(v-2i\pi\right)}{\sinh \frac{v}{2}} \ e^{-\frac{\beta\left(v-\frac{2i\pi (\beta-\tau)}{\beta}\right)^2}{4
		\kappa \tau(\beta-\tau) }}.
\end{equation}
It is convenient to center the gaussian weight in \eqref{case2} $v=0$ through the shift $v\rightarrow v+\frac{2i\pi (\beta-\tau)}{\beta}$;
\begin{equation} 
\label{case3}
\begin{split}
\langle\mathcal{O}^{(\frac12)}(\tau)\rangle_\beta^{\rm disk}=-\mbox{\small$ \frac{i\kappa^{-\frac12} \beta^{\frac32}}{8 \pi^{\frac32} \tau ^{3/2}
			(\beta -\tau )^{3/2}}$} \int_{-\infty}^{+\infty} \mathrm{d}v \mbox{\small$\left(v^2+\frac{2i \pi\left(\beta-2\tau \right)}{\beta}v+\frac{4\pi^2 \tau \left(\beta-\tau\right)}{\beta^2}\right)$} \frac{e^{-\frac{\beta v^2}{4
			\kappa \tau(\beta-\tau) }}}{{\sinh}\left(\frac{v}{2}-\frac{i \pi \tau}{\beta}\right)}
\end{split}
\end{equation}
Now we replace $1/\sinh(\cdots)$ with its representation in terms of exponentials
\begin{equation} 
\label{case4}
\text{csch}\left(\frac{v}{2}-\frac{\pi i\tau}{\beta}\right)=
\frac{2}{v}e^{-\frac{\pi i\tau}{\beta}}\left(\frac{v ~e^{\frac{v}{2}}}{e^{{v}-\frac{2\pi i\tau}{\beta}}-1}\right),
\end{equation}
and recognize that the quantity between parenthesis is the generating functional of the so-called generalized Apostol-Bernoulli polynomials of degree one. The definition of these polynomials for general degree and some of their properties are briefly diskussed in app. \ref{BernApp}. Therefore we directly write
\begin{equation} 
\label{case5}
\text{csch}\left(\frac{v}{2}-\frac{\pi i\tau}{\beta}\right)=2e^{-\frac{\pi i\tau}{\beta}}\sum_{n=0}^\infty \mathcal{B}^{(1)}_n \left(\frac12,e^{-\frac{2\pi i\tau}{\beta}}\right)\frac{v^{n-1}}{n!}.
\end{equation}
If we integrate in $v$ term by term using the expansion \eqref{case5}, we encounter only powers of $v$ averaged over a gaussian weight. We find convenient to treat separately the even and the odd powers in  \eqref{case5}. Reordering the powers in $\kappa$ produced by the gaussian integrations, we obtain the following perturbative series for the expectation value of the bi-local operator with $\lambda=\frac12:$
\begin{equation} 
\begin{split}
\langle\mathcal{O}^{(\frac12)}(\tau)\rangle_\beta^{\rm disk}=&\mbox{\small$ \frac{e^{-\frac{\pi i\tau}{\beta}}}{ \pi^{\frac12} 
		}$} \sum_{p=0}^\infty \mbox{\small$\frac{2^{2p}\kappa^{p}\tau^{p}\left(\beta-\tau\right)^{p} \Gamma \left[p+\frac12\right]}{\left(2p\right)!\beta^{p}}$} \ \left[\mbox{\small$\frac{\left(\beta-2\tau \right)}{\tau \left(\beta-\tau\right)}$} \mathcal{B}^{(1)}_{2p} \left(\frac12,e^{-\frac{2\pi i\tau}{\beta}}\right)-\right. \\ &\left.-\mbox{\small$\frac{i }{\left(2p+1\right)}$} \mathcal{B}^{(1)}_{2p+1} \left(\frac12,e^{-\frac{2\pi i\tau}{\beta}}\right)
 -\mbox{\small$\frac{ip \beta}{ \pi \tau \left(\beta-\tau\right)}$} \mathcal{B}^{(1)}_{2p-1} \left(\frac12,e^{-\frac{2\pi i\tau}{\beta}}\right)\right].
\end{split}
\end{equation}
The case $\lambda=1$ is slightly more involved: the explicit form of the residue is
\begin{equation} 
\label{case1n=1}
\mathrm{Res}_{u=v}\left(\frac{u \ e^{-\frac{u^2}{4
				\kappa(\beta-\tau)}}}{(\cosh\frac{v}{2}-\cosh\frac{u}{2})^2 }\right)=-\frac{2 e^{-\frac{v^2}{4\kappa(\beta-\tau)}}}{\sinh^2\frac{v}{2}} \left( \frac{v^2}{\kappa(\beta-\tau)}+v \coth \left(\frac{v}{2}\right)-2\right),
\end{equation}
and we can perform again the previous analysis. This time the dependence on $1/\sinh(\cdots)$ is accompanied by higher powers, namely $1/\sinh^2(\cdots)$ and $1/\sinh^3(\cdots)$. With computations similar to the case $\lambda=1/2$, we can also obtain with little effort 
the all order expansion 
\begin{equation} 
\begin{split}
&\langle\mathcal{O}^{(1)}(\tau)\rangle_\beta^{\rm disk}=\mbox{\small$\frac{e^{-\frac{2i\pi \tau}{\beta}}}{\pi^{\frac12}}$} \ \sum_{p=0}^\infty \mbox{\small$\frac{2^{2p}\kappa^{p}\tau^{p}\left(\beta-\tau\right)^{p}\Gamma\left(p+\frac12\right)}{\left(2p\right)!\beta^{p}}$} \ \left[\left(\mbox{\footnotesize$\frac{p-1}{2p-1}\frac{\beta^2}{\tau^2 \left(\beta-\tau\right)^2}-\frac{3}{\tau\left(\beta-\tau\right)}$}\right)\mathcal{B}^{(2)}_{2p}\left(1,e^{-\frac{2\pi i\tau}{\beta}}\right)\right. \\
&\left.-\mbox{\small$\frac{3i\pi}{\left(2p+1\right)}\frac{\beta-2\tau}{\beta\tau\left(\beta-\tau\right)}$}\mathcal{B}^{(2)}_{2p+1}\left(1,e^{-\frac{2\pi i\tau}{\beta}}\right)-\mbox{\small$\frac{4 \pi^2}{\beta^2\left(2p+1\right)\left(2p+2\right)}$}\mathcal{B}^{(2)}_{2p+2}\left(1,e^{-\frac{2\pi i\tau}{\beta}}\right)\right. \\
& \left.-\mbox{\footnotesize$\frac{ip}{2\pi}\frac{\beta\left(\beta-2\tau\right)}{\tau^2 \left(\beta-\tau\right)^2}$}\mathcal{B}^{(2)}_{2p-1}\left(1,e^{-\frac{2\pi i\tau}{\beta}}\right)\right] 
\end{split}
\end{equation}
where also generalized Apostol-Bernoulli polynomials of degree 2 appears in the expansion.

The same analysis can be easily carried out for the trumpet. Here the starting point is
\begin{equation} 
\begin{split}
\langle\mathcal{O}^{(\lambda)}(\tau)\rangle_\beta^{\rm tr.}=-\mbox{\small$\frac{ \pi^2\Gamma \left(2\lambda\right) \mathcal{N}}{2^{2\lambda+1}\kappa^2 \left(\beta-\tau\right)^{\frac12}\tau^{\frac32}}$} \int_{-\infty}^{+\infty} \mathrm{d}v \ \mathrm{Res}_{u=v}\left(\frac{u \ e^{-\frac{u^2}{4
			\kappa(\beta-\tau)}}}{\left(\cosh\frac{v}{2}-\cosh\frac{u}{2}\right)^{2\lambda}} \right) \ e^{-\frac{\left(v+2b \pi \right)^2}{4
		\kappa\tau }}
\end{split}
\end{equation} 
 The case $\lambda=1/2$ and $\lambda=1$ are again obtained along the same lines diskussed above and one gets
\begin{equation} 
\begin{split}
\langle\mathcal{O}^{(\frac12)}(\tau)\rangle_\beta^{\rm tr.}=\mbox{\small$\frac{e^{-\frac{b\pi \tau}{\beta}} }{ \pi^{\frac12}}$} \sum_{p=0}^\infty \mbox{\small$\frac{2^{2p}\kappa^{p}\tau^{p}\left(\beta-\tau\right)^{p}\Gamma \left[p+\frac12\right]}{\left(2p!\right)\beta^{p}}$} \ \left(\mbox{\small$\frac{1}{\tau}$}\mathcal{B}^{(1)}_{2p} \left(\frac12,e^{-\frac{2b\pi \tau}{\beta}}\right)\mbox{-\small$\frac{2b\pi}{\left(2p+1\right)\beta}$}\mathcal{B}^{(1)}_{2p+1} \left(\frac12,e^{-\frac{2b\pi \tau}{\beta}}\right)\right)
\end{split}
\end{equation} 
and
\begin{equation} 
\begin{split}
&\langle\mathcal{O}^{(1)}(\tau)\rangle_\beta^{\rm tr.}=\mbox{\small$\frac{e^{-\frac{2b\pi \tau}{\beta}} }{ \pi^{\frac12}}$} \sum_{p=0}^\infty \mbox{\small$\frac{2^{2p}\kappa^{p}\tau^{p}\left(\beta-\tau\right)^{p}\Gamma \left[p+\frac12\right]}{\left(2p!\right)\beta^{p}}$} \ \left[\left(\mbox{\small$\frac{1}{\tau^2}-\frac{\beta+b\pi\tau}{ \tau^2 \left(\beta-\tau\right)\left(2p-1\right)}$}\right)\mathcal{B}^{(2)}_{2p} \left(1,e^{-\frac{2b\pi \tau}{\beta}}\right)\right. \\
&\left. -\mbox{\small$\frac{4b\pi}{ \beta \tau \left(2p+1\right)}$} \mathcal{B}^{(2)}_{2p+1}(1,e^{-\frac{2\pi b\tau}{\beta}})\!+\!\mbox{\small$\frac{4 b^2 \pi^2}{\beta^2\left(2p+1\right)\left(2p+2\right)}$}\mathcal{B}^{(2)}_{2p+2} \left(1,e^{-\frac{2b\pi \tau}{\beta}}\right)\!+\!\mbox{\small$\frac{p\beta}{ \tau^2 \left(\beta-\tau\right)\left(2p-1\right)}$}\mathcal{B}^{(2)}_{2p-1}\left(1,e^{-\frac{2\pi b\tau}{\beta}}\right) \right.\\
&\left.+\mbox{\small$\frac{1}{ \tau^2 \left(\beta-\tau\right)\left(2p-1\right)}$}\left(\beta\mathcal{B}^{(3)}_{2p}\left(1,e^{-\frac{2\pi b\tau}{\beta}}\right) -\mbox{\small$\frac{2b\pi\tau}{\left(2p+1\right)}$}\mathcal{B}^{(3)}_{2p+1}\left(1,e^{-\frac{2\pi b\tau}{\beta}}\right)\right)\right]
\end{split}
\end{equation}
The trumpet expressions become a little bit more cumbersome since we lose the symmetry $\tau\to \beta-\tau$, as expected since we are working in the zero winding sector. As a consequence, we also notice the presence of generalized Apostol-Bernoulli polynomials of degree 3.
\subsection{The case of generic n}
\label{genericn}
Having trained with the simplest cases, we are ready now to perform the computation for generic semi-integer $\lambda$. In app. \ref{BernApp} we have shown that the residue in \eqref{residue} can always be computed in terms of { generalized} Apostol-Bernoulli polynomials. The structure of the answer is
\be
 \mathrm{Res}_{u=v}\left(\frac{u \ e^{-\alpha u^2}}{\left(\cosh\frac{v}{2}-\cosh\frac{u}{2}\right)^{2\lambda}} \right)=e^{-\alpha v^2}f(v)
\ee
where $\alpha=\frac{1}{4\kappa(\beta-\tau)}$ and
\be
\label{Res1}
f(v)=e^{\lambda v}
 \sum_{\ell=0}^{2\lambda-1} \frac{(-2)^{4\lambda-\ell-1}{\alpha^{\frac{2\lambda-\ell-2}{2}}}\mathcal{H}_{2\lambda-\ell}\left(\sqrt{\alpha} v\right)}{(2\lambda-\ell-1)!}
 \sum_{j=0}^\ell \frac{\Bb^{(2\lambda)}_{\ell-j}\left(\lambda\right)
 \mathcal{B}^{(2\lambda)}_{j+2\lambda}\left(\lambda,e^{v}\right)}{(\ell-j)!(2\lambda+j)!}.
\ee
In \eqref{Res1} $\mathcal{H}_n(x)$ stands for the usual Hermite polynomials, while $\mathcal{B}^n_k(x, y)$ and $\Bb^n_k(x)$ are respectively
 {generalized} Apostol-Bernoulli of degree $n$ and { generalized} Bernoulli polynomials. Their definition and some of their properties are diskussed in app. \ref{BernApp}. Then the remaining integral in $v$ takes the following form:
\begin{equation}
\begin{split}
\langle\mathcal{O}^{(\lambda)}(\tau)\rangle_\beta^{\rm disk}=\mbox{\small$- \frac{i\pi^2 \Gamma(2\lambda)\mathcal{N}_d}{ 2^{2\lambda+2}\kappa ^3 \tau ^{3/2}
			(\beta -\tau )^{3/2}}$} \int_{-\infty}^{+\infty} \mathrm{d}v \ (v-2i\pi) f(v)\ e^{-\frac{(v-2 i \pi )^2}{4
			\kappa \tau }-\frac{v^2}{4
			\kappa(\beta-\tau)}}.
\end{split}
\end{equation} 
Next we perform the shift $v\mapsto v+\frac{2\pi i(\beta-\tau)}{\beta}$ to move the gaussian center around $v=0$. We find
\begin{equation}
\label{uuvv}
\begin{split}
\langle\mathcal{O}^{(\lambda)}(\tau)\rangle_\beta^{\rm disk}=\mbox{\small$- \frac{i\pi^2 \Gamma(2\lambda)\kappa^{2\lambda-\frac32} \beta^{\frac32}}{ 2^{2\lambda+3}\pi^{\frac72} \tau ^{\frac32}
			(\beta -\tau )^{\frac32}}$} \int_{-\infty}^{+\infty} \mathrm{d}v \ \left( v-\frac{2\pi i\tau}{\beta}\right) f\left( v+\frac{2\pi i(\beta-\tau)}{\beta}\right)\ e^{-\frac{\beta v^2}{4
			\kappa\tau (\beta-\tau)}}.
\end{split}
\end{equation} 
Using the result \eqref{fundexpans}, we can immediately expand $f\left( v+\frac{2\pi i(\beta-\tau)}{\beta}\right)$ in powers of $v$
\begin{align}
\label{expaf}
f\biggl( v+&\left.\frac{2\pi i(\beta-\tau)}{\beta}\right)=\\
 &=(-1)^{2\lambda}\sum_{m=0}^\infty \frac{v^m}{m!}
 \sum_{\ell=0}^{2\lambda-1} \frac{(-2)^{4\lambda-\ell-1}{\alpha^{\frac{2\lambda-\ell-2}{2}}}\mathcal{H}_{2\lambda-\ell}\left(\sqrt{\alpha}\left( v+\frac{2\pi i(\beta-\tau)}{\beta}\right)\right)}{(2\lambda-\ell-1)!} c^{(\lambda)}_{\ell,m}(\beta,\tau),\nn
\end{align}
where $c^{(\lambda)}_{\ell,m}(\beta,\tau)$ denotes the following combination
\be
c^{(\lambda)}_{\ell,m}(\beta,\tau)=e^{-\frac{2\pi i\lambda\tau}{\beta}}\sum_{j=0}^\ell \frac{(j+m)!~}{(\ell-j)!(2\lambda+j+m)!j!}~\Bb^{(2\lambda)}_{\ell-j}\left(\lambda\right)
 \B^{(2\lambda)}_{j+2\lambda+m}\left(\lambda;e^{-\frac{2\pi i\tau}{\beta}}\right).
\ee
Plugging the expansion \eqref{expaf} into the integral \eqref{uuvv}, we get a series representation for our correlator
\begin{align}
\label{uuvv1}
\langle\mathcal{O}^{(\lambda)}&(\tau)\rangle_\beta^{\rm disk}=\mbox{\small$- \frac{i\pi^2 (-1)^{2\lambda}\Gamma(2\lambda)\kappa^{2\lambda-\frac32} \beta^{\frac32}}{ 2^{2\lambda+3}\pi^{\frac72} \tau ^{\frac32}
			(\beta -\tau )^{\frac32}}$}
			\sum_{m=0}^\infty \frac{1}{m!}
 \sum_{\ell=0}^{2\lambda-1} \frac{(-2)^{4\lambda-\ell-1}}{(2\lambda-\ell-1)!} c^{(\lambda)}_{\ell,m}(\beta,\tau)\times\\
&\times			\alpha^{\frac{2\lambda-\ell-2}{2}} \int_{-\infty}^{+\infty} \mathrm{d}v~v^m \mathcal{H}_{2\lambda-\ell}\left(\sqrt{\alpha}\left( v+\frac{2\pi i(\beta-\tau)}{\beta}\right)\right)
			 \ \left( v-\frac{2\pi i\tau}{\beta}\right)\ e^{-\frac{\beta v^2}{4
			\kappa\tau (\beta-\tau)}}.\nn
\end{align}
To single out the dependence on $\kappa$, we scale our variable of integration as follows $v\to 2\sqrt{\frac{\kappa\tau(\beta-\tau)}\beta}$ and we get
\begin{align}
		\langle\mathcal{O}^{(\lambda)}&(\tau)\rangle_\beta^{\rm disk}	=\mbox{\small$- \frac{i\pi^2 (-1)^{2\lambda}\Gamma(2\lambda)\kappa^{2\lambda-\frac32} \beta^{\frac32}}{ 2^{2\lambda+3}\pi^{\frac72} \tau ^{\frac32}
			(\beta -\tau )^{\frac32}}$}
			\sum_{m=0}^\infty \picco{\frac{2^{m+1} (\kappa\tau(\beta-\tau))^{\frac{m+1}2}}{\beta^{\frac{m+1}2}m!}}
 \sum_{\ell=0}^{2\lambda-1}\picco{ \frac{(-2)^{4\lambda-\ell-1}}{(2\lambda-\ell-1)!} c^{(\lambda)}_{\ell,m}(\beta,\tau)}\times\nn\\
&\!\!\!\times\!\!	\picco{		\left(\frac1{4\kappa(\beta-\tau)}\right)^{\frac{2\lambda-\ell-2}{2}} }\!\!\!\int_{-\infty}^{+\infty} \!\!\!\!\!\mathrm{d}v~v^m \mathcal{H}_{2\lambda-\ell}\picco{\left(\sqrt{\frac\tau\beta} v+\frac{i \pi \sqrt{\beta -\tau }}{\beta \sqrt{\kappa }}\right)}
			 \picco{\left(2\sqrt{\frac{\kappa\tau(\beta-\tau)}{\beta}} v-\frac{2\pi i\tau}{\beta}\right)}\ e^{-v^2}.
\end{align}
Next we exploit a simple rule holding for Hermite polynomials with shifted argument
\be
\mathcal{H}_{2\lambda-\ell}\picco{\left(\sqrt{\frac\tau\beta} v+\frac{i \pi \sqrt{\beta -\tau }}{\beta \sqrt{\kappa }}\right)}=
\sum_{k=0}^{2\lambda-\ell}\binom{2\lambda-\ell}{k}\mathcal{H}_{k}\picco{\left(\sqrt{\frac\tau\beta} v\right)\left(\frac{2\pi i \sqrt{\beta -\tau }}{\beta \sqrt{\kappa }}\right)^{2\lambda-\ell-k}},
\ee
to rearrange our correlator in the form
\begin{align}	&\langle\mathcal{O}^{(\lambda)}(\tau)\rangle_\beta^{\rm disk}	=
\mbox{\small$- \frac{i\pi^2 (-1)^{2\lambda}\Gamma(2\lambda)\kappa^{2\lambda-\frac32} \beta^{\frac32}}{ 2^{2\lambda+3}\pi^{\frac72} \tau ^{\frac32}
			(\beta -\tau )^{\frac32}}$}
			\sum_{m=0}^\infty \picco{\frac{2^{m+1} (\kappa\tau(\beta-\tau))^{\frac{m+1}2}}{\beta^{\frac{m+1}2}m!}}
 \sum_{\ell=0}^{2\lambda-1}\picco{ \frac{(-2)^{4\lambda-\ell-1}}{(2\lambda-\ell-1)!} c^{(\lambda)}_{\ell,m}(\beta,\tau)}\times\nn\\
&\times\!\!		\picco{\left(\frac1{4\kappa(\beta-\tau)}\right)^{\frac{2\lambda-\ell-2}{2}} }\sum_{k=0}^{2\lambda-\ell}\picco{
\binom{2\lambda-\ell}{k}\picco{\left(\frac{2\pi i \sqrt{\beta -\tau }}{\beta \sqrt{\kappa }}\right)^{2\lambda-\ell-k}}}
			\! \picco{\left(\!\sqrt{\frac{4\kappa\tau(\beta-\tau)}{\beta}} \PP_{k,m+1}-\frac{2\pi i\tau}{\beta}\PP_{k,m}\!\right)},
\end{align}
where
\begin{equation}
\label{Pkm}
\PP_{k,m}=\int_{-\infty}^\infty dv~ v^m \mathcal{H}_{k}\left(\sqrt{\frac\tau\beta} v\right) e^{-v^2}. 
\end{equation}
This integral yields a polynomial of order $k$ in $\sqrt{\frac\tau\beta}$ and its explicit expression in terms of the associated Legendre function is given in \eqref{associated}.
Obviously $\PP_{k,m}$ is different from zero only when $m+k$ is an even number. We shall use this selection rule to rearrange the two  contributions proportional to $ \PP_{k,m+1}$ [(A)]  and $ \PP_{k,m}$ [(B)] respectively. For the former the selection rule is $m+k+1=2 p$ with $p=1,\cdots,\infty$.
We can use this result to replace the sum over $m$ with a sum over $p$ 
\begin{align}
&(A)=\mbox{\small$- \frac{i\pi^2 (-1)^{2\lambda}\Gamma(2\lambda)}{ 2^{2\lambda}\pi^{\frac72} 
			}$}
			\sum_{p=1}^\infty \picco{\frac{\kappa^{p+\ell} }{(2p-k-1)!} \left(\frac{4\kappa\tau(\beta-\tau)}{\beta}\right)^{p+\ell-2\lambda}}\times\\
&\times\!\! \sum_{\ell=0}^{2\lambda-1}\sum_{k=0}^{\mathrm{min}(2\lambda-\ell,2 p-1)}\!\!\!\!\!\!\!\!\picco{ \frac{(-2)^{4\lambda-\ell-1}}{(2\lambda-\ell-1)!} c^{(\lambda)}_{\ell,2p-k-1}(\beta,\tau)}\picco{
\binom{2\lambda-\ell}{k}\picco{\left(\frac{4\pi i\sqrt{\tau} (\beta -\tau)}{\beta^\frac32 }\right)^{2\lambda-\ell-k}}}\picco{\left(\frac\tau\beta\right)^{\frac{2\lambda-\ell-2}{2}} }
		\!\!\!\!\!\!\!\!	 \picco{ \PP_{k,2p-k}}.\nn
\end{align}
Next we introduce a new index $n\equiv k+\ell$, which simply counts the power of the coupling constant
\begin{align}
(A)&=\mbox{\small$\frac{ (-\pi i)^{2\lambda+1}\Gamma(2\lambda)}{ (2\beta)^{2\lambda}\pi^{\frac52} 
			}$}\sum_{n=1}^\infty \picco{\left(\frac{4\kappa\tau(\beta-\tau)}{\beta}\right)^{n} }\sum_{\ell=0}^{\mathrm{min}(n,2\lambda-1)}\frac{(-2)^{4\lambda-\ell-1}}{(2\lambda-\ell-1)!}\times\\
			&\times\sum_{k=0}^{\mathrm{min}(2\lambda-\ell,2 n-2\ell-1)}\!\!
\picco{ \frac{c^{(\lambda)}_{\ell,2n-2\ell-k-1}(\beta,\tau)}{(2n-2\ell-k-1)!}} \picco{
\binom{2\lambda-\ell}{k}}\picco{\left(\frac{\beta^2 }{4\pi i \tau(\beta -\tau)}\right)^{\ell+k}}\picco{\left(\frac\tau\beta\right)^{\frac{k}{2}-1} }
			 \picco{ \PP_{k,2n-2\ell-k}}.\nn
			\end{align}
The same analysis is done for the latter contribution, taking into account the constraint $m+k=2 p$ with $p=0,1,\cdots,\infty$:
\begin{align}
&(B)=\mbox{\small$\left(\frac{2\pi i\tau}\beta\right)\frac{i\pi^2 (-1)^{2\lambda}\Gamma(2\lambda)}{ 2^{2\lambda}\pi^{\frac72} 
			}$}
			\sum_{p=0}^\infty \picco{\frac{\kappa^{p+\ell} }{(2p-k)!} \left(\frac{4\kappa\tau(\beta-\tau)}{\beta}\right)^{p+\ell-2\lambda}}\times\\
&\times \sum_{\ell=0}^{2\lambda-1}\sum_{k=0}^{\mathrm{min}(2p,2\lambda-\ell)}
 \picco{ \frac{(-2)^{4\lambda-\ell-1}}{(2\lambda-\ell-1)!} c^{(\lambda)}_{\ell,2p-k}(\beta,\tau)}\picco{
\binom{2\lambda-\ell}{k}\picco{\left(\frac{4\pi i\sqrt{\tau} (\beta -\tau)}{\beta^\frac32 }\right)^{2\lambda-\ell-k}}}\picco{\left(\frac\tau\beta\right)^{\frac{2\lambda-\ell-2}{2}} }\!\!\!
			 \picco{ \PP_{k,2p-k}}.\nn
\end{align}
Setting again $n=p+\ell$, we get
\begin{align}
&(B)=\mbox{\small$-\left(\frac{2\pi i\tau}\beta\right)\frac{ (-\pi i)^{2\lambda+1}\Gamma(2\lambda)}{ (2\beta)^{2\lambda}\pi^{\frac52} 
			}$}
			\sum_{n=0}^\infty \picco{ \left(\frac{4\kappa\tau(\beta-\tau)}{\beta}\right)^{n}} \times\\
&\times\!\!\!\!\!\! \!\!\sum_{\ell=0}^{\mathrm{min}{(n,2\lambda-1)}}\sum_{k=0}^{\mathrm{min}(2n-2\ell,2\lambda-\ell)}\!\!\!
 \picco{ \frac{(-2)^{4\lambda-\ell-1} c^{(\lambda)}_{\ell,2n-2\ell-k}(\beta,\tau)}{(2n-2\ell-k)!(2\lambda-\ell-1)!}}\picco{
\binom{2\lambda-\ell}{k}}\!\picco{\left(\frac{\beta^2}{4\pi i \tau (\beta -\tau)}\right)^{\ell+k}}\!\!
\picco{\left(\frac\tau\beta\right)^{\frac{k}{2}-1} }
			\!\!\!\!\!\!\! \picco{ \PP_{k,2n-2\ell-k}}.\nn
\end{align}
Combining the two contributions, we obtain the final expansion
\begin{align}
\langle&\mathcal{O}^{(\lambda)}(\tau)\rangle_\beta^{\rm disk}=\picco{\frac{\pi^{2\lambda}}{\beta^{2\lambda}\sin^{2\lambda}\frac{\pi\tau}{\beta}}}
+
\mbox{\small$\frac{ (-\pi i)^{2\lambda+1}\Gamma(2\lambda)}{ (2\beta)^{2\lambda}\pi^{\frac52} 
			}$}\sum_{n=1}^\infty \picco{\left(\frac{4\kappa\tau(\beta-\tau)}{\beta}\right)^{n} }~\picco{\sum_{\ell=0}^{\mathrm{min}(n,2\lambda-1)}}\!\!\frac{(-2)^{4\lambda-\ell-1}}{(2\lambda-\ell-1)!}\times\nn\\ &\times\left[\picco{\sum_{k=0}^{\mathrm{min}(2 n-2\ell-1,2\lambda-\ell)}}
\picco{ \frac{c^{(\lambda)}_{\ell,2n-2\ell-k-1}(\beta,\tau)}{(2n-2\ell-k-1)!}} \nn\right.		
\picco{
\binom{2\lambda-\ell}{k}}\picco{\left(\frac{\beta^2 }{4\pi i \tau(\beta -\tau)}\right)^{\ell+k}}\picco{\left(\frac\tau\beta\right)^{\frac{k}{2}-1} }
			 \picco{ \PP_{k,2n-2\ell-k}}-\nn\\
			 &-\picco{{2\pi i}}\sum_{k=0}^{\mathrm{min}(2n-2\ell,2\lambda-\ell)}
 \picco{ \frac{c^{(\lambda)}_{\ell,2n-2\ell-k}(\beta,\tau)}{(2n-2\ell-k)!}}\picco{
\binom{2\lambda-\ell}{k}}\picco{\left(\frac{\beta^2}{4\pi i \tau (\beta -\tau)}\right)^{\ell+k}}
\picco{\left(\frac\tau\beta\right)^{\frac{k}{2}} }
			 \picco{ \PP_{k,2n-2\ell-k}}\Biggr].
			\end{align}
We make a couple of observations on the above expression: first of all, we notice that at sufficiently large order in $\kappa$ the non-trigonometric dependence on $\tau$ cannot grow arbitrarily, being a polynomial bounded by the weight of the bi-local operators itself. Moreover we expect the original symmetry $\tau\to\beta-\tau$ to be preserved by the expansion: looking at the structure of the coefficients it is not manifest but we checked its presence till the order $\kappa^6$. Making explicit this symmetry should probably simplify the final formula. A second remark concerns the trigonometric dependence of the generic perturbative term and its singularity properties as
 $\tau\to 0$. The trigonometric dependence is completely encoded into the coefficients $c^{(\lambda)}_{\ell,m}(\beta,\tau)$: we expect the presence of negative powers of $\sin(\tau/\beta),$ generating a singular behaviour at small $\tau$. This fact is also evident from the singularity appearing in this limit for the generalized Apostol-Bernoulli polynomials.
\section{Expansion for $\beta\to\infty$}
It is now interesting to concentrate on the zero temperature limit of the bi-local correlator to check explicitly the agreement with \cite{Mertens:2020pfe}.
The structure of our integrals simplifies significantly as $\beta\to\infty$: moreover we observe that both the disk and the trumpet share the same behaviour in this regime, since  we expected that as the total boundary length diverges (while keeping $b$ fixed) the presence of an hole in the interior becomes negligible. 

However this limit cannot be directly extracted from the final result of subsec. \ref{genericn} since the limit of {generalized} Apostol-Bernoulli polynomial $\mathcal{B}^{(\ell)}_n(x,\mu)$ is discontinuous when $\mu$ approaches one. Therefore it is convenient to go back to eq. \eqref{bi-localinte2} and take the limit $\beta\to\infty$ at this level. The limit of the integrand and of
its normalization is smooth and we get
\begin{equation}
\langle\mathcal{O}^{(\lambda)}(\tau)\rangle_{\beta \rightarrow \infty}=\frac{\kappa^{2\lambda}\Gamma(2\lambda)}{2^{2\lambda+4} 
 \pi^{\frac52}\kappa ^{\frac32} \tau
  ^{\frac32} }
 \int_{-\infty}^\infty du dv~ 
\frac{(u-2\pi i) (v-2\pi i)}{(\cosh \frac{u}{2}+\cosh \frac{v}{2})^{2\lambda}} e^{-\frac{(v-2 i \pi )^2}{4 \kappa \tau }}
  \end{equation}
The integral over $u$ can be now evaluated in closed form. The linear term in $u$ vanishes since it is odd, while the contribution proportional to $2\pi i$ yields
\begin{equation}
\langle\mathcal{O}^{(\lambda)}(\tau)\rangle_{\beta \rightarrow \infty}=\frac{8\pi i\kappa^{2\lambda}\Gamma(2\lambda)}{2^{2\lambda+4} 
 \pi^{\frac52}\kappa ^{\frac32} \tau
  ^{\frac32} }
 \int_{-\infty}^\infty dv~ 
\frac{ (v-2\pi i)}{\sinh^{2\lambda}\frac{v}2} e^{-\frac{(v-2 i \pi )^2}{4 \kappa \tau }}\mathcal{Q}_{2\lambda-1}\left(\coth\frac{v}2\right),
  \end{equation}
where we have used that 
\begin{equation}
\label{LegendreQ}
\frac{\mathcal{Q}_{n}(\coth\frac{v}2)}{\sinh^{n+1}\frac{v}2}=\frac1{4}\int_{-\infty}^{+\infty}\frac{\mathrm{d}u}{\left(\cosh\frac{v}{2}+ \cosh \frac{u}2\right)^{n+1}}.
\end{equation}
In eq. \eqref{LegendreQ} $\mathcal{Q}_{n}(z)$ stands for the so-called Legendre function of the second kind. Next we eliminate the dependence on the linear factor $(v-2\pi i)$ by integrating by parts and we write
\begin{equation}
\langle\mathcal{O}^{(\lambda)}(\tau)\rangle_{\beta \rightarrow \infty}=\frac{16\pi i\kappa^{2\lambda}\Gamma(2\lambda)\kappa \tau}{2^{2\lambda+4} 
 \pi^{\frac52}\kappa ^{\frac32} \tau
  ^{\frac32} }
 \int_{-\infty}^\infty dv~ 
 e^{-\frac{(v-2 i \pi )^2}{4 \kappa \tau }}
\frac{d}{dv}\left(\frac{ \mathcal{Q}_{2\lambda-1}\left(\coth\frac{v}2\right)}{\sinh^{2\lambda}\frac{v}2}\right) ,
  \end{equation}
Exploiting the recurrence relation for the Legendre function of the second kind and its derivatives for integer indices, it is straightforward to show that
\begin{equation}
\frac{\mathrm{d}}{\mathrm{d}v}\left[\frac{ \mathcal{Q}_{2\lambda-1}\left(\coth\frac{v}2\right)}{\sinh^{2\lambda}\frac{v}2}\right]=-\lambda\frac{ \mathcal{Q}_{2\lambda}\left(\coth\frac{v}2\right)}{\sinh^{2\lambda}\frac{v}2}
\end{equation}
Thus
\begin{equation}\label{inte}
\langle\mathcal{O}^{(\lambda)}(\tau)\rangle_{\beta \rightarrow \infty}=-\frac{16\pi i\lambda\kappa^{2\lambda}\Gamma(2\lambda)\kappa \tau}{2^{2\lambda+4} 
 \pi^{\frac52}\kappa ^{\frac32} \tau
  ^{\frac32} }
 \int_{-\infty}^\infty dv~ 
 e^{-\frac{(v-2 i \pi )^2}{4 \kappa \tau }}
\frac{ \mathcal{Q}_{2\lambda}\left(\coth\frac{v}2\right)}{\sinh^{2\lambda}\frac{v}2}.
\end{equation}
Although the structure of the integrand suggests the possible presence of a singularity at $v=0$, it is not difficult to check this 
singularity is only apparent. In fact a careful analysis of the integrand shows that it is completely regular at $v=0$.
When $2\lambda$ is an integer $\mathcal{Q}_{2\lambda}$ can be expressed in terms of the Legendre polynomials. Specifically the
following identity holds
\begin{equation}
\mathcal{Q}_{2\lambda}\left(\mbox{\small$\coth \frac{v}{2}$}\right)=\frac12 \mathcal{P}_{2\lambda} \left(\mbox{\small$\coth \frac{v}{2}$}\right) v-\mathcal{W}_{2\lambda-1}\left(\mbox{\small$\coth \frac{v}{2}$}\right) \quad \end{equation}
with
\be
\mathcal{W}_{2\lambda-1}\left(\mbox{\small$\coth \frac{v}{2}$}\right)=\sum_{k=1}^{2\lambda}\frac{1}{k}\mathcal{P}_{k-1}\left(\mbox{\small$\coth \frac{v}{2}$}\right) \mathcal{P}_{2\lambda-k}
\left(\mbox{\small$\coth \frac{v}{2}$}\right)
\ee
Therefore the Legendre function of the second kind is not periodic under the shift $v\rightarrow v+2i\pi$, but we have
\begin{equation}
\mathcal{Q}_{2\lambda}\left(\mbox{\small$\coth \frac{v}{2}$}\right)\rightarrow \mathcal{Q}_{2\lambda}\left(\mbox{\small$\coth \frac{v}{2}$}\right)+i\pi \mathcal{P}_{2\lambda}\left(\mbox{\small$\coth \frac{v}{2}$}\right).
\end{equation}
If we perform this shift in our integral we find 
\begin{equation}\label{inte2}
\langle\mathcal{O}^{(\lambda)}(\tau)\rangle_{\beta \rightarrow \infty}=-\frac{16\pi i(-1)^{2\lambda}\lambda\kappa^{2\lambda}\Gamma(2\lambda)\kappa \tau}{2^{2\lambda+4} 
 \pi^{\frac52}\kappa ^{\frac32} \tau
  ^{\frac32} }
 \int_{-\infty}^\infty dv~ 
 e^{-\frac{v^2}{4 \kappa \tau }}
\frac{\left[ \mathcal{Q}_{2\lambda}\left(\coth\frac{v}2\right)+\pi i \mathcal{P}_{2\lambda}\left(\mbox{\small$\coth \frac{v}{2}$}\right)\right]}{\sinh^{2\lambda}\frac{v}2}.
\end{equation}
The combination $\sinh^{-2\lambda}\left( \frac{v}{2}\right)\mathcal{Q}_{2\lambda}\left(\coth \frac{v}{2}\right)$ is an odd function and so its contribution to the integral \eqref{inte2} identically vanishes. So we are left with the term proportional to $\mathcal{P}_{2\lambda}$ only, i.e.
\begin{equation}\label{inte3}
\langle\mathcal{O}^{(\lambda)}(\tau)\rangle_{\beta \rightarrow \infty}=\frac{16\pi^2(-1)^{2\lambda}\lambda\kappa^{2\lambda}\Gamma(2\lambda)\kappa \tau}{2^{2\lambda+4} 
 \pi^{\frac52}\kappa ^{\frac32} \tau
  ^{\frac32} }
 \int_{-\infty}^\infty dv~ 
 e^{-\frac{v^2}{4 \kappa \tau }}
\frac{ \mathcal{P}_{2\lambda}\left(\mbox{\small$\coth \frac{v}{2}$}\right)}{\sinh^{2\lambda}\frac{v}2}.
\end{equation}
A remark is now in order. The final integral is singular at $v=0$. If we perform, as we should, the translation $v\rightarrow v+2i\pi$ as a change of path in
the complex $v-$plane, we have to deform a little bit the contour to avoid precisely $v=0$ since the integrand possesses a pole there.
This small deformation provides us the prescription on how to regularize the singularity (it is  PV-like prescription). In the following,
we neglect this issue and regularize this singularity using an analytic regularization, which is more straightforward and produces the same result. 

Next we use the following representation of the Legendre polynomials $\mathcal{P}_n(x)$
\begin{equation}
\mathcal{P}_n (x)=\frac{1}{2^{n}}\sum_{k=0}^{n}\binom{n}{k}^2 \left(x-1\right)^{n} \left(\frac{x+1}{x-1}\right)^{k}
\end{equation}
For $x=\coth\frac{v}{2}$ this representation simplifies
\begin{equation}
\mathcal{P}_{2\lambda} \left(\coth \frac{v}{2}\right)= \left(\frac{1}{e^v-1}\right)^{2\lambda}\sum_{k=0}^{2\lambda}\binom{2\lambda}{k}^2 \ e^{kv}
\end{equation}
and our integral becomes
\begin{equation}\label{gi2}
\langle\mathcal{O}^{(\lambda)}(\tau)\rangle_{\beta \rightarrow \infty}=\frac{16\pi^2(-1)^{2\lambda}\lambda\kappa^{2\lambda}\Gamma(2\lambda)\kappa \tau}{2^{2\lambda+4} 
 \pi^{\frac52}\kappa ^{\frac32} \tau
  ^{\frac32} } \sum_{k=0}^{2\lambda}\mbox{\small$\binom{2\lambda}{k}^2$}
 \int_{-\infty}^\infty dv~ 
 e^{-\frac{v^2}{4 \kappa \tau }}
\left(\frac{1}{e^v-1}\right)^{4\lambda} \ e^{v \left(k+\lambda\right)}.
\end{equation}
We can now expand part of the integrand in terms of generalized Bernoulli polynomials
\begin{equation}
\label{Laur1}
\left(\frac{1}{e^v-1}\right)^{4\lambda} \ e^{v \left(k+\lambda\right)}=\sum_{n=0}^{\infty} \frac{\mathrm{B}_{n}^{(4\lambda)}\left(k+\lambda\right)}{n!} \ v^{n-4\lambda}.
\end{equation}
This expansion explicitly exhibits the aforementioned poles present in the integrand.  Equation \eqref{Laur1} is a Laurent series which contains negative powers up to
$-4\lambda$.
Then we have to compute
\begin{equation}\label{gi3}
\begin{split}
\langle\mathcal{O}^{(\lambda)}(\tau)\rangle_{\beta \rightarrow \infty}=&\frac{16\pi^2(-1)^{2\lambda}\lambda\kappa^{2\lambda}\Gamma(2\lambda)\kappa \tau}{2^{2\lambda+4} 
 \pi^{\frac52}\kappa ^{\frac32} \tau
  ^{\frac32} } \sum_{k=0}^{2\lambda}\mbox{\small$\binom{2\lambda}{k}^2$}
\sum_{n=0}^{\infty} \frac{\mathrm{B}_{n}^{(4\lambda)}\left(k+\lambda\right)}{n!} \int_{-\infty}^\infty \ \mathrm{d}v \
v^{n-4\lambda}
 e^{-\frac{v^2}{4 \kappa \tau }}.
\end{split}
\end{equation}
Since $4\lambda$ is even, the integral is different from zero only for even $n$. If we set $n=2p$ with $p\in\mathds{N}$, the gaussian integral can be now easily performed and we always get
\begin{equation}
\int_{-\infty}^\infty \ \mathrm{d}v \
v^{2\left(p-2\lambda\right)} e^{-\frac{v^2}{4 \kappa \tau }}=(2\sqrt{\kappa\tau})^{2(p-2\lambda)+1}\Gamma\left[\frac{2p-4\lambda+1}{2}\right],
\end{equation}
where we have defined the integral for negative powers of $v$ by analytic continuation.
By substituting it in \eqref{gi3} we find
\begin{equation}\label{gi5}
\begin{split}
\langle\mathcal{O}^{(\lambda)}(\tau)\rangle_{\beta \rightarrow \infty}=&\frac{(2\lambda)!(-1)^{2\lambda}}{2^{2\lambda} 
 \pi^{\frac12} } \frac{1}{\tau^{2\lambda}}
\sum_{p=0}^{\infty} \frac{(4\kappa\tau)^p}{(2p)!}\sum_{k=0}^{2\lambda}\mbox{\small$\binom{2\lambda}{k}^2$}\mathrm{B}_{2p}^{(4\lambda)} (k+\lambda)\Gamma\left[\frac{2p-4\lambda+1}{2}\right].
\end{split}
\end{equation}
To better understand the structure of this
perturbative expansion it is convenient to separate positive and negative powers of $\tau$. Recalling the value of the Gamma function
for seminteger values of the argument, we immediately find
\begin{equation}\label{beta}
\begin{split}
\langle\mathcal{O}^{(\lambda)}(\tau)\rangle_{\beta \rightarrow \infty}
&= \frac{\mbox{\footnotesize$\left(2\lambda\right) !$}}{\tau^{2\lambda}} \left\{\sum_{p=0}^{2\lambda-1}\frac{\left(\kappa \tau\right)^{p}}{\mbox{\footnotesize$\left(2p\right) !$}} \mbox{\small$(-1)^{p} \ \frac{\left(2\lambda-p\right)!}{\left(4\lambda-2p\right)!}$}
\sum_{k=0}^{2\lambda} \mbox{\small$\binom{2\lambda}{k}^2$} \mathrm{B}_{2p}^{(4\lambda)}\left(k+\lambda\right)+\right. \\
&+\left.\mbox{\small$\frac{(-1)^{2\lambda}}{2^{2\lambda}}$}\sum_{r=2\lambda}^{\infty}\frac{\left(2\kappa \tau\right)^{r}}{\mbox{\footnotesize$\left(2r\right) !$}}\mbox{\small$\left(2r-4\lambda-1\right)!!$} 
\sum_{k=0}^{2\lambda} \mbox{\small$\binom{2\lambda}{k}^2$} \mathrm{B}_{2r}^{(4\lambda)}\left(k+\lambda\right) \right\}
\end{split}
\end{equation}
One can easily check that for $\lambda=\frac12$ and $\lambda=1$ this result exactly reproduces the expressions given by \cite{Mertens:2020pfe}, where the first perturbative orders are presented.
\section{Mordell integrals  and bi-local correlators for integer $2\lambda$} 
As anticipated in the introduction, in this section our goal is to show that the bi-local correlator can be in general expressed as a combination of Mordell integrals.
Concretely, let us go back to the case $2\lambda\in\mathds{N}$ and to be more specific we focus on $\lambda=1/2$ for the disk.  The expression \eqref{case3} can be rewritten as follows
\begin{equation} 
\begin{split}
\langle\mathcal{O}^{(\frac12)}(\tau)\rangle_\beta^{\rm disk}=-\mbox{\small$ \frac{2\pi^\frac32 i\kappa^{-\frac12} \beta^{\frac32}e^{\frac{i \pi  \tau }{\beta }}}{  \tau ^{3/2}
			(\beta -\tau )^{3/2}}$} \int_{-\infty}^{+\infty} \mathrm{d}v \mbox{\small$\left( \underset{ \genfrac{}{}{0pt}{1}{\phantom{c}}{(a)}  }{v^2}+\underset{(b)}{\frac{i \left(\beta-2\tau \right)}{\beta}v}+\underset{(c)}{\frac{ \tau \left(\beta-\tau\right)}{\beta^2}}\right)$}  \frac{e^{-\frac{\pi^2\beta v^2}{
			\kappa \tau(\beta-\tau) }+{\pi v}}}{e^{2\pi v}-e^{\frac{2 \pi  i\tau }{\beta }}},
\end{split}
\end{equation} 
where we have also scaled the integration variable by $2\pi$.  We recognize three different contributions: the third one can be immediately identified with the so-called Mordell integral \cite{mordell1933}, which appears in number theory and in the theory of Mock-theta functions \cite{2008arXiv0807.4834Z}.  The general form of the Mordell integral is 
\begin{equation}
\mathcal{M}(x,\theta,\omega)=
\int_{-\infty}^\infty dt~ \frac{e^{\pi i\omega t^2-2\pi x t}}{e^{2\pi t}-e^{2\pi i\theta}}=e^{-\pi i(\theta^2 \omega+2\theta x+2\theta)}
\frac{F[(x+\theta)/\omega,-1/\omega]+ i\omega F[x+\theta \omega,\omega]}{\omega \theta_{11}(x+\theta\omega,\omega)}.
\end{equation}
The function $F(x,\omega)$ admits a $q-$expansion  of the form
\begin{equation}
F[x,\omega]=-i\sum_{m\in\mathds{Z}}\frac{(-1)^m q^{(m+1/2)^2}e^{2\pi i (m+1/2)x}}{1+q^{2m+1}},
\end{equation}
where $q=e^{\pi i\omega}$. The denominator is one of the usual Jacobi theta function and its $q-$expansion  is
\begin{equation}
\theta_{11}(x,\omega)=-i\sum_{m\in\mathds{Z}}{(-1)^m q^{(m+1/2)^2}e^{2\pi i (m+1/2)x}}.
\end{equation}
In our case, we have $x=-\frac{1}{2}$, $\theta=\frac{\tau}{\beta}$ and $\omega=\frac{\pi i \beta}{
			\kappa \tau(\beta-\tau) }$. The other two contributions, (a) and (b), are proportional to the second and first derivatives of the Mordell integral with respect to $x$.  Then the  complete result for the bi-local correlator at $\lambda=1/2$:
			\begin{equation} 
\begin{split}
\langle\mathcal{O}^{(\frac12)}&(\tau)\rangle_\beta^{\rm disk}=- \frac{2\pi^\frac32 i\kappa^{-\frac12} \beta^{\frac32}e^{\frac{i \pi  \tau }{\beta }}}{  \tau ^{3/2}
			(\beta -\tau )^{3/2}}
			\left( 
			\frac{1}{4\pi^2}\partial_x^2\mathcal{M}\left(-\frac{1}{2},\frac{\tau}{\beta},\frac{\pi i \beta}{
			\kappa \tau(\beta-\tau)}\right)-\right.\\
&\left.-			{\frac{i \left(\beta-2\tau \right)}{2\pi\beta}}
			\partial_x\mathcal{M}\left(-\frac{1}{2},\frac{\tau}{\beta},\frac{\pi i \beta}{
			\kappa \tau(\beta-\tau)}\right)
			+{\frac{ \tau \left(\beta-\tau\right)}{\beta^2}}\mathcal{M}\left(-\frac{1}{2},\frac{\tau}{\beta},\frac{\pi i \beta}{
			\kappa \tau(\beta-\tau)}\right)\right) \end{split}
\end{equation} 
The structure of  the bi-local correlator for $2\lambda$ generic integer is not so different. In fact, by carefully  inspecting   \eqref{uuvv}
we can easily verify that it is given by a sum of integrals of the following form
\begin{equation}
\int_{-\infty}^\infty dt~t^n \frac{e^{\pi i\omega t^2-2\pi x t}}{(e^{2\pi t}-e^{2\pi i\theta})^m},
\end{equation}
where $m$ and $n$ are integers. However any integral of this kind can be evaluated in terms of the  original Mordell integral:
\begin{equation}
\int_{-\infty}^\infty dt~t^n \frac{e^{\pi i\omega t^2-2\pi x t}}{(e^{2\pi t}-e^{2\pi i\theta})^m}=\frac{(-1)^n}{(m-1)!}\left (\frac{1}{2\pi}\partial_x\right)^n \left(\frac{e^{-2\pi i\theta}}{2\pi i}\partial_\theta\right)^{m-1}\mathcal{M}(x,\theta,\omega)
\end{equation}
In other words, the correlators are  completely controlled by this kind of functions.
\section{Final comments and outlook}
In this work, we have considered JT bi-local correlators of operators with positive weight $\lambda$, on the disk and the trumpet topologies. The perturbative series associated to these correlation functions is harder to obtain than in the parent case $\lambda \in -{\mathbb N}/2$, recently studied in \cite{Mertens:2020pfe} in the zero temperature limit. We have been able nevertheless to distill some aspects of the $\kappa$ expansion of the two-point function, checking the agreement of the exact non-perturbative expression with the Schwarzian perturbation theory for any value of $\beta$. In the particular case of $\lambda \in {\mathbb N}/2$, we derived an all-order formula for the perturbative contributions, that becomes particularly handful in the limit of infinite $\beta$. We have also shown that the exact expression for our bi-local correlators is closely related to the Mordell integral, a basic constituent in the theory of Mock-modular forms \cite{chen}. 

There are some lessons that we can draw from our computations and some directions that could be worth to study further.
A feature that we may explore from the knowledge of the entire perturbative series is, for example, the nature of possible non-perturbative contributions to the full answer. For instance,
let us consider the   coefficient  $c_r$ in the case $\beta\to\infty$.  We can easily read it from \eqref{beta}:
\begin{equation}
\begin{split}
c_r=\frac{\left(2\lambda\right) ! \left(2r-4\lambda-1\right)!!}{2^{2\lambda}}\frac{2^r \tau^{r-2\lambda}}{\mbox{\footnotesize$\left(2r\right) !$}}\left[ \left(\sum_{k=0}^{2\lambda} \mbox{\small$\binom{2\lambda}{k}^2$} \mathcal{B}_{2r}^{(4\lambda)}\left(k+\lambda\right)\right) \  \right]
\end{split}
\end{equation}
To understand its behaviour for large value of $r$ we need to know the behaviour of the generalized Bernoulli polynomials in that limit.
This aspect was diskussed in detail in \cite{LOPEZ2010197}, where it was found that the dominant contribution is
\begin{equation}
\label{asy1}
\begin{split}
\mathcal{B}_{2r}^{m}(z)\simeq - \left(2r\right)! \left[\beta_{1}^{m}\frac{e^{2\pi i z}}{\left(2\pi i\right)^{2r}}+\beta_{-1}^{m}\frac{e^{-2\pi i z}}{\left(-2\pi i\right)^{2r}}\right]
\end{split}
\end{equation}
where $ \displaystyle
\beta_{k}^{m}\left(n,z\right)\simeq \frac{\left(-1\right)^{m-1}n^{m-1}}{\left(m-1\right)!}.
$
At leading order these coefficients are independent of  $k$ and \eqref{asy1} collapses to
\begin{equation}
\begin{split}
\mathcal{B}_{2r}^{m}(z)\simeq &
\frac{\left(2r\right)! \left(-1\right)^{m+r}\left(2r\right)^{m-1}2}{\left(2\pi\right)^{2r}\left(m-1\right)!}\cos \left(2\pi z\right).
\end{split}
\end{equation}
If we choose $m=4\lambda$ the coefficient $c_r$ for $r\to\infty$ takes the form
\begin{equation}
\begin{split}
c_r\simeq &\frac{\left(2\lambda\right) ! \left(2r-4\lambda-1\right)!!}{2^{2\lambda}}\frac{2^r \tau^{r-2\lambda}}{\left(2r\right) !}\frac{\left(2r\right)! \left(-1\right)^{4\lambda+r}\left(2r\right)^{4\lambda-1}2}{\left(2\pi\right)^{2r}\left(4\lambda-1\right)!} (-1)^{2\lambda}\sum_{k=0}^{2\lambda} \binom{2\lambda}{k}^2=\\
=&\frac{\left(2\lambda\right) ! \left(2r-4\lambda-1\right)!!}{2^{2\lambda}}\frac{2^r \tau^{r-2\lambda}}{\left(2r\right) !}\frac{\left(2r\right)! \left(-1\right)^{4\lambda+r}\left(2r\right)^{4\lambda-1}2}{\left(2\pi\right)^{2r}\left(4\lambda-1\right)!} (-1)^{2\lambda}\frac{4\lambda !}{\left(2\lambda!\right)^2},
\end{split}
\end{equation}
where we have performed the sum over the square of the binomial coefficients.
We can now easily complete our large  $r-$expansion with the help of the Stirling formula. After some tedious algebra we find
\begin{equation}
\begin{split}
c_r
&=\frac{4\lambda\left(-1\right)^{6\lambda} \tau^{r-2\lambda}e^{2\lambda}}{\sqrt{\pi}\left(2\lambda\right)!} \  \frac{\left(-1\right)^{r}r^{2\lambda-\frac32}r!}{\pi^{2r}}
\end{split}
\end{equation}
The coefficient grows as a power  times $r!$ and its global sign alternates with the parity of $r$. Thus  the perturbative series appears to be Borel-summable: in fact the leading pole appearing  in the Borel-transform is located on the negative axis and thus one could argue that non-perturbative instanton-like configurations should not play any role here. On the one hand, there is no guarantee that the Borel resummation of a Borel summable series reconstructs the non-perturbative answer. There are sufficient conditions for this to be the case, which typically require strong analyticity conditions on the underlying non-perturbative function. On the other hand, in most of the examples of Borel summable series in quantum theories, Borel resummation does reconstruct the correct answer (see \cite{Grassi:2014cla} for a lucid diskussion of these topics). The actual determination of non-perturbative configurations, if any, remains therefore an important issue for future investigations, as well as to understand their possible physical meaning with respect to the boundary gravitons appearing in Schwarzian perturbation theory. 

The obvious extension of the present work would consist in studying the perturbative series associated to general four-point correlators of bi-local operators. While the exact form on the disk and the trumpet is well known \cite{Mertens:2017mtv,Iliesiu:2019xuh}, much less has been learned on its perturbative incarnation, due to the appearing in the out-of-time-ordered case of a very complicated vertex function inside the integrals. The relevant 6-$j$ symbols involved there can be expressed through Wilson function and, in principle, one could try to perform an expansion using the analytical structure of the full amplitude. The success of such computation would certainly improve our understanding of  the properties of the associated gravitational S-matrix. 

Another generalization of our investigations would concern the perturbative aspects of two-point functions in presence of defects \cite{Mertens:2019tcm}. The trumpet correlators studied here are just a particular example within this class, being associated to a bi-local operator with the insertion of a hyperbolic defect in the bulk and computed without taking into account the winding sectors  \cite{Mertens:2019tcm}. It could be interesting to extend our analysis to the winding case and to consider elliptic and parabolic defects too. The fate and the physics of bi-local correlators in presence of multiple defects \cite{Maxfield:2020ale} or for deformed JT gravity \cite{Witten:2020wvy} could be also explored. It would be nice also to understand the character of perturbative contributions to bi-local correlators from boundary fluctuations in higher-genus geometry \cite{Blommaert:2020seb}.

Finally we point out that the correlators studied here were obtained in \cite{Mertens:2020hbs} from boundary correlators of minimal Liouville string, exploiting a particular double-scaling limit. It would be interesting to see if the Mordell structure, underlying the exact form of the bi-local correlator on the disk, could be understood from a Liouville perspective.

\bigskip

\paragraph{Acknowledgements:} We thank Marisa Bonini and Itamar Yaakov for several discussion on different aspects of this paper. This work has been supported in part by Italian Ministero dell’Istruzione, Universit\`a e Ricerca (MIUR), and Istituto Nazionale di Fisica Nucleare (INFN) through the “Gauge and String Theory” (GAST)  research project.
 \newpage
\appendix
{\large \bf Appendices}
\addcontentsline{toc}{section}{\protect\textbf{Appendices}}
\section{Evaluation of the Residue}
\label{BernApp}
Most of our results can be expressed in terms of the  so-called { generalized}   Apostol-Bernoulli polynomials $\B^{(\ell)}_n(x;\mu)$. If $\ell\in\mathds{N}$, they are defined through the generating function:
 \begin{equation}
 \label{appa1}
\left(\frac{t}{\mu e^t-1}\right)^\ell e^{x t}=\sum_{n=\ell}^\infty\B^{(\ell)}_n(x;\mu)\frac{t^n}{n!}.
\end{equation}
This definition implies that $\B^{(\ell)}_n(x;\mu)=0$ for $n=0,\dots,\ell-1$.  The { generalized}   Apostol-Bernoulli numbers $\B^{(\ell)}_n(\mu)$ are then given by
\be
 \label{appa2}
\B^{(\ell)}_n(\mu)\equiv  \B^{(\ell)}_n(0;\mu).
\ee
The familiar Bernoulli polynomials are recovered when we set $\ell=1$ and $\mu=1$.  The  explicit form of this polynomials can be obtained as follows. First  we consider the combination $\left(\frac{t}{\mu e^t-1}\right)^\ell$ and write its formal expansion in power of $e^t-1$.
\begin{align}
 \label{appa3}
\left(\frac{t}{\mu e^t-1}\right)^\ell =&\frac{t^\ell}{(\mu-1)^\ell} \left(\frac{\mu-1}{\mu e^t-1}\right)^\ell  =\frac{t^\ell}{(\mu-1)^\ell} \left(1+\frac{\mu}{\mu-1}(e^t-1)\right)^{-\ell} =\nn\\
=&t^\ell\sum_{k=0}^\infty \binom{k+\ell-1}{k} \frac{(-\mu)^k}{(\mu-1)^{k+\ell}}(e^t-1)^k
\end{align}
Next we use that
\be
 \label{appa4}
(e^t-1)^k=k!\sum_{r=k}^\infty S(r,k) \frac{t^r}{r!}
\ee
where $S(r,k)$ denotes the Stirling numbers of the second kind. Thus
\begin{align}
 \label{appa5}
\left(\frac{t}{\mu e^t-1}\right)^\ell =&\sum_{k=0}^\infty \sum_{r=k}^\infty \binom{k+\ell-1}{k} \frac{ k! (-\mu)^k}{(\mu-1)^{k+\ell}} S(r,k) \frac{t^{r+\ell}}{r!}=\nn\\
=&\sum_{r=0}^\infty  \frac{t^{r+\ell}}{r!}\sum_{k=0}^r \binom{k+\ell-1}{k} \frac{ k! (-\mu)^k}{(\mu-1)^{k+\ell}} S(r,k)=\nn\\
=&\sum_{r=0}^\infty  \frac{t^{r+\ell}}{(r+l)!} \ell!\sum_{k=0}^r \binom{r+\ell}{r}\binom{k+\ell-1}{k} \frac{ k! (-\mu)^k}{(\mu-1)^{k+\ell}} S(r,k)
\end{align}
From eq. \eqref{appa5} we can immediately extract a representation for the {generalized}   Apostol-Bernoulli numbers $\B^{(\ell)}_n(\mu)$  by setting $r=n-\ell$.
\begin{align}
 \label{appa6}
\B^{(\ell)}_n(\mu)=&\ell!\sum_{k=0}^{n-\ell} \binom{n}{n-\ell}\binom{k+\ell-1}{k} \frac{ k! (-\mu)^k}{(\mu-1)^{k+\ell}} S(n-\ell,k)=\nn\\
=&\ell!\sum_{k=0}^{n-\ell} \binom{n}{\ell}\binom{k+\ell-1}{k} \frac{ k! (-\mu)^k}{(\mu-1)^{k+\ell}} S(n-\ell,k).
\end{align}
Given the {generalized}   Apostol-Bernoulli numbers $\B^{(\ell)}_n(\mu)$ , it is easy to write down the expansion for the polynomial
\be
 \label{appa7}
\B^{(\ell)}_n(x;\mu)=\sum_{k=0}^n\binom{n}{k} \B_{n-k}^{(\ell)}(\mu) x^k.
\ee
The case $\mu=1$ are simply known as   {generalized}   Bernoulli polynomials  and we shall denote them as $\Bb^{(\ell)}_n(x)$.  Obviously we can also introduce the
 { generalized}   Bernoulli numbers, $\Bb^{(\ell)}_n\equiv\Bb^{(\ell)}_n(0)$.  These polynomials are not simply obtained by taking the limit for $\mu\to 1 $ of the previous explicit expressions. The latter are in fact divergent in this limit. 
 
\medskip

In  the following we show that the residue appearing in the computation of the bi-local correlator when $2\lambda$ is an integer can be expressed in terms of these generalized quantities. We start by observing that
\begin{align}
 \label{appa8}
\frac{1}{\left(\cosh\frac{v}{2}-\cosh\frac{u}{2}\right)^{n}}=&
-(-2)^{n-1}e^{\frac{1}{2} n v}  \left(\frac{2}{y}\right)^{2n} \underset{\rm (A)}{\frac{\left(\frac{y}{2}\right)^n e^{\frac{1}{4} n y}}{\left(e^{\frac{y}{2}}-1\right)^{n}} }
\underset{\rm (B)}{ \frac{\left(\frac{y}{2}\right)^n   e^{\frac{1}{4} n y} }{  \left(e^{v+\frac{y}{2}}-1\right)^{n}}}=\nn\\
 =&-(-2)^{n-1} e^{\frac{1}{2} n v} \left(\frac{2}{y}\right)^{n}\sum_{\ell=0}^\infty \left(\frac{y}{2}\right)^{\ell}\sum_{j=0}^\ell \frac{\Bb^{(n)}_{\ell-j}\left(\frac{n}{2}\right)
 \B^{(n)}_{j+n}\left(\frac{n}{2};e^{v}\right)}{(\ell-j)!(n+j)!},
   \end{align}
where we have introduced $y=u-v$ to keep  a compact notation.  We have expanded the factor (A) in terms of   { generalized}   Bernoulli polynomials, while
the remaining factor (B) has been expressed as series whose coefficients are the  {generalized}   Apostol-Bernoulli polynomials for $\lambda=e^v$.  If we use the property
\be
 \mathcal{B}^{(n)}_{k}(n-x,\mu)=\frac{(-1)^k}{\mu^n} \mathcal{B}^{(n)}_{k}(x,{\mu^{-1}})
\ee
we find that 
\be
 \Bb^{(n)}_{k}\left(\frac n2\right)=(-1)^k  \Bb^{(n)}_{k}\left(\frac n2\right)\ \ \ \ \ \ \   e^{\frac{1}{2} n v} \mathcal{B}^{(n)}_{k}\left(\frac n2,e^v\right)={(-1)^k} e^{-\frac{1}{2} n v} \mathcal{B}^{(n)}_{k}\left(\frac n2,e^{-v}\right).
\ee
The first identity implies that $ \Bb^{(n)}_{k}\left(\frac n2\right)$ vanishes for odd $k$.
Next we observe that 
the combination $u \exp(-\alpha u^2)$ can be written as
\begin{align}
 \label{appa9}
u e^{-\alpha u^2}=\frac{1}{2}
\sum_{n=0}^\infty (-2)^n\frac{\alpha^{\frac{n-1}{2}}}{n!}\mathcal{H}_{n+1}(\sqrt{\alpha} v)e^{-\alpha v^2}\left(\frac{y}{2}\right)^n
\end{align}
where $\mathcal{H}_n(u)$ stand for the usual Hermite polynomials. Thus
 we find the following Laurent expansion for the function $f(u)=\frac{u e^{-\alpha u^2}}{\left(\cosh\frac{v}{2}-\cosh\frac{u}{2}\right)^{n}}$:
\begin{align}
 \label{appa10}
f(u)=&-(-2)^{n-1}  \left(\frac{2}{y}\right)^{n}  e^{-\alpha v^2+\frac{n}{2}v}\sum_{p=0}^\infty \left(\frac{y}{2}\right)^p e^{-\alpha v^2}\times\nn\\
&\times \sum_{\ell=0}^p(-2)^{p-\ell}\frac{\alpha^{\frac{p-\ell-1}{2}}}{(p-\ell)!}\mathcal{H}_{p-\ell+1}(\sqrt{\alpha} v) \sum_{j=0}^\ell \frac{\Bb^{(n)}_{\ell-j}\left(\frac{n}{2}\right)
 \B^{(n)}_{j+n}\left(\frac{n}{2};e^{v}\right)}{(\ell-j)!(n+j)!}.
 \end{align}
It is a trivial exercise to extract the relevant residue form \eqref{appa10}. We get
\begin{align}
 \label{appa11}
\mathrm{Res}[f(u)]_{u=v}=&e^{-\alpha v^2+\frac{n}{2}v}
 \sum_{\ell=0}^{n-1} \frac{(-2)^{2 n-\ell-1}{\alpha^{\frac{n-\ell-2}{2}}}\mathcal{H}_{n-\ell}(\sqrt{\alpha} v)}{(n-\ell-1)!}
 \sum_{j=0}^\ell \frac{\Bb^{(n)}_{\ell-j}\left(\frac{n}{2}\right)
 \B^{(n)}_{j+n}\left(\frac{n}{2};e^{v}\right)}{(\ell-j)!(n+j)!}.
\end{align}
\section{Some useful expansion for generalized Apostol-Bernoulli Polynomials}
In ref \cite{LUO2005290,pippo} they provide  the following expansion for the {  generalized} Apostol-Bernoulli polynomials in terms
of the  {generalized} Bernoulli polynomials (i.e. $\mu=1$):
\be
\B_j^{(n)}(x,\mu)=e^{-x\log\mu}\sum_{k=0}^\infty \binom{j+k-n}{k}{\binom{j+k}{k}}^{-1} \Bb^{(n)}_{k+j}(x)
\frac{(\log\mu)^k}{k!}
\ee
This expansion suggests that it is possible  to expand the {  generalized} Apostol-Bernoulli polynomials at a given 
$\mu=\mu_1\mu_2$ in terms of the same polynomials at $\mu=\mu_1$. In fact, exploiting the properties of logarithms
\begin{align}
&\B_j^{(n)}(x,\mu_1\mu_2)=e^{-x(\log\mu_1+\log\mu_2)}\sum_{k=0}^\infty \binom{j+k-n}{k}{\binom{j+k}{k}}^{-1} \Bb^{(n)}_{k+j}(x)
\frac{(\log\mu_1+\log\mu_2)^k}{k!}=\nn\\
=&e^{-x(\log\mu_1+\log\mu_2)}\sum_{k=0}^\infty \sum_{\ell=0}^k\binom{j+k-n}{k}{\binom{j+k}{k}}^{-1} \Bb^{(n)}_{k+j}(x)
\frac{1}{k!}\binom{k}{\ell}(\log\mu_1)^\ell (\log\mu_2)^{k-\ell}
\end{align}
We can disentangle the two sums by setting $k=\ell+m$. Then the two sums becomes independent:
\begin{align}
=&e^{-x(\log\mu_1+\log\mu_2)}\sum_{m=0}^\infty \sum_{\ell=0}^\infty\binom{j+m+\ell-n}{m+\ell}{\binom{j+m+\ell}{m+\ell}}^{-1} \Bb^{(n)}_{m+\ell+j}(x)
\frac{(\log\mu_1)^\ell}{\ell !} \frac{(\log\mu_2)^{m}}{m!}.
\end{align}
Let use rearrange the binomials coefficient as follows and perform the sum over $\ell$:
\begin{align}
=&e^{-x(\log\mu_1+\log\mu_2)}\sum_{m=0}^\infty \sum_{\ell=0}^\infty\frac{\binom{j+m-n}{m} \binom{j+\ell+m-n}{\ell}}{\binom{j+m}{m} \binom{j+l+m}{\ell}}
\Bb^{(n)}_{m+\ell+j}(x)
\frac{(\log\mu_1)^\ell}{\ell !} \frac{(\log\mu_2)^{m}}{m!}=\nn\\
=&e^{-x \log\mu_2}\sum_{m=0}^\infty \frac{\binom{j+m-n}{m}}{\binom{j+m}{m}}
\B^{(n)}_{m+j}(x,\mu_1) \frac{(\log\mu_2)^{m}}{m!}.
\end{align}
Therefore we have shown
\be
\B_j^{(n)}(x,\mu_1\mu_2)=e^{-x \log\mu_2}\sum_{m=0}^\infty \frac{\binom{j+m-n}{m}}{\binom{j+m}{m}}
\B^{(n)}_{m+j}(x,\mu_1) \frac{(\log\mu_2)^{m}}{m!}.
\ee
If we apply this result to our specific case we get
\begin{align}
\label{fundexpans}
e^\frac{nv}{2}\B^{(n)}_{j+n}\left(\frac{n}{2};e^{v-\frac{2\pi i\tau}{\beta}}\right)=\sum_{m=0}^\infty \frac{\binom{j+m}{m}}{\binom{j+m+n}{m}}
\B^{(n)}_{m+j}\left(\frac{n}{2};e^{-\frac{2\pi i\tau}{\beta}}\right) \frac{v^{m}}{m!}.
\end{align}
\section{Computing gaussian integrals of Hermite polynomials}
We consider the following integral
\begin{equation}
I_k=\int_{-\infty}^\infty dv~  \mathcal{H}_{k}\left(a v\right) e^{-v^2+x v}.
\end{equation}
To find its expression for general $k$ we construct the following generating functional
\begin{align}
G(t)=&\sum_{k=0}^\infty  \frac{t^k}{k!}I_k=\int_{-\infty}^\infty dv~  e^{-v^2+x v+2 a t v-t^2}=\sqrt{\pi } e^{\left(a^2-1\right) t^2+a t x+\frac{x^2}{4}}=\nn\\
=&\sqrt{\pi } e^{-\left(\sqrt{1-a^2}t\right)^2 +2\frac{a}{\sqrt{1-a^2}}(\sqrt{1-a^2} t) \frac{x}{2}+\frac{x^2}{4}}=\sqrt{\pi}e^{\frac{x^2}{4}}\sum_{k=0}^\infty 
\frac{t^k (1-a^2)^{\frac{k}{2}}}{k!} \mathcal{H}_k\left(\frac{a x}{2\sqrt{1-a^2}}\right),
\end{align}
where  we used 
\be
\sum_{n=0}^\infty \frac{t^n}{n!}\mathcal{H}_n(a v)=e^{2 a t v-t^2}.
\ee
Thus
\be
I_k= \sqrt{\pi}(1-a^2)^{\frac{k}{2}}\mathcal{H}_k\left(\frac{a x}{2\sqrt{1-a^2}}\right)e^{\frac{x^2}{4}}.
\ee
The integral $P_{km}$ defined in \eqref{Pkm} can be computed by taking the $m^{th}$ derivative with respect to $x$ of $I_k$ and then setting $x=0$
\begin{align}
\PP_{km}=&\sqrt{\pi}\left.\partial_x^m I_k\right|_{x=0}=\left.\sum_{j=0}^m  \binom{m}{j}  (1-a^2)^{\frac{k}{2}}
\partial_x^j \mathcal{H}_k\left(\frac{a x}{2\sqrt{1-a^2}}\right)\partial^{m-j}_x e^{\frac{x^2}{4}}\right|_{x=0}=\nn\\
=&\sqrt{\pi}\left.\sum_{j=0}^m  \binom{m}{j}  
\frac{k! a^j \left(1-a^2\right)^{\frac{k-j}2} H_{k-j}\left(\frac{a x}{2 \sqrt{1-a^2}}\right)}{(k-j)!}e^{\frac{x^2}{4}} \left(-\frac{i}{2}\right)^{m-j} H_{m-j}\left(\frac{i x}{2}\right)\right|_{x=0}=\nn\\
=&\pi^\frac32\sum_{j=0}^m  \binom{m}{j}  
\frac{(-i)^{m-j}\Gamma(k+1) a^j \left(1-a^2\right)^{\frac{k-j}2}  2^{k-j}}{\Gamma(k-j+1)\Gamma \left(\frac{1}{2} (j-k+1)\right)\Gamma \left(\frac{1}{2} (j-m+1)\right)}.
\end{align}
This sum can be easily evaluated in terms of hypergeometric functions once we extend the range of $j$ to infinity.  In fact  the 
generic term, once written in terms of $\Gamma-$function, vanishes for $j\ge m+1$. We find
\begin{align}
\label{associated}
\PP_{km}=&\pi ^{3/2} 2^k (-i)^m \left(1-a^2\right)^{\frac{k}{2}} \left(\frac{\,
   _2F_1\left(-\frac{k}{2},-\frac{m}{2};\frac{1}{2};\frac{a^2}{a^2-1}\right)}{\Gamma \left(\frac{1-k}{2}\right) \Gamma
   \left(\frac{1-m}{2}\right)}+\frac{2 i a \,
   _2F_1\left(\frac{1-k}{2},\frac{1-m}{2};\frac{3}{2};\frac{a^2}{a^2-1}\right)}{\sqrt{1-a^2} \Gamma \left(-\frac{k}{2}\right) \Gamma
   \left(-\frac{m}{2}\right)}\right)=\nn\\
   =& (-i)^m  2^{\frac{k-m-1}{2}} \pi\left(1-a^2\right)^{\frac{k-m-1}{4}} \mathcal{P}_{\frac{1}{2}
   (k-m-1)}^{\frac{1}{2} (k+m+1)}\left(-\frac{i a}{\sqrt{1-a^2}}\right),
\end{align}
where $\mathcal{P}^\mu_\nu(x)$ is the Associated Legendre Function.
\bibliographystyle{nb}
\bibliography{PTJT}

\end{document}